\documentclass[11pt]{article}

\usepackage[margin=1.0in]{geometry}

\usepackage{graphics}
\usepackage{url}
\usepackage{caption}
\usepackage{amsmath}
\usepackage{hyperref}
\usepackage{cite}
\usepackage{slashed}
\usepackage[colorinlistoftodos, shadow]{todonotes}
\usepackage{titlesec}
\usepackage{feynarts}
\usepackage{gensymb}
\usepackage{ulem}

\def\refeq#1{\mbox{(\ref{#1})}}
\def\reffi#1{\mbox{Fig.~\ref{#1}}}
\def\reffis#1{\mbox{Figs.~\ref{#1}}}
\def\refta#1{\mbox{Table~\ref{#1}}}

\def\refse#1{\mbox{Section~\ref{#1}}}

\def\citere#1{\mbox{Ref.~\cite{#1}}}
\def\citeres#1{\mbox{Refs.~\cite{#1}}}

\def\al{\alpha}

\def\ga{\gamma}
\def\de{\delta}

\def\la{\lambda}
\def\si{\sigma}

\def\La{\Lambda}

\newcommand{\GeV}{\unskip\,\mathrm{GeV}}
\newcommand{\MeV}{\unskip\,\mathrm{MeV}}
\newcommand{\TeV}{\unskip\,\mathrm{TeV}}
\newcommand{\fba}{\unskip\,\mathrm{fb}}

\def\mathswitch#1{\relax\ifmmode#1\else$#1$\fi}
\def\mathswitchr#1{\relax\ifmmode{\mathrm{#1}}\else$\mathrm{#1}$\fi}
\def\mathswitchit#1{\relax\ifmmode{#1}\else$#1$\fi}

\newcommand{\Pu}{u}
\newcommand{\Pd}{d}
\newcommand{\Ps}{s}
\newcommand{\Pc}{c}
\newcommand{\Pt}{\mathswitchr t}
\newcommand{\Pp}{\mathswitchr p}

\newcommand{\Pq}{\mathswitchit q}
\newcommand{\Pl}{\mathswitch l}
\newcommand{\Plp}{\mathswitch l^+}
\newcommand{\Pe}{\mathswitchr e}

\newcommand{\PWp}{\mathswitchr {W^+}}
\newcommand{\PWm}{\mathswitchr {W^-}}

\newcommand{\PW}{\mathswitchr W}
\newcommand{\PZ}{\mathswitchr Z}

\newcommand{\Pg}{g}
\newcommand{\PH}{\mathswitchr H}

\newcommand{\Pnl}{\mathswitch {\nu_\Pl}}

\newcommand{\MW}{\mathswitch {M_\PW}}

\newcommand{\MZ}{\mathswitch {M_\PZ}}

\newcommand{\GF}{\mathswitch {G_\mu}}

\renewcommand{\O}{{\cal O}}

\newcommand{\ri}{{\mathrm{i}}}

\newcommand{\rd}{{\mathrm{d}}}

\newcommand{\M}{{\cal {M}}}
\renewcommand{\L}{{\cal L}}

\newcommand{\cut}{{\mathrm{cut}}}
\newcommand{\jet}{{\mathrm{jet}}}

\newcommand{\EW}{{\mathrm{EW}}}
\newcommand{\QCD}{{\mathrm{QCD}}}

\newcommand{\OS}{{\mathrm{OS}}}

\newcommand{\sub}{{\mathrm{sub}}}

\newcommand{\veto}{{\mathrm{veto}}}

\newcommand{\NLO}{{\mathrm{NLO}}}

\newcommand{\alphas}{\alpha_{s}}
\newcommand{\abs}[1]{\left|#1\right|}

\newcommand{\klam}[1]{\left(#1\right)}

\newcommand{\md}{\mathrm{d}}
\newcommand{\mr}[1]{\mathrm{#1}}
\newcommand{\MtW}{M_{\mathrm{T},\Plp\nu}}
\newcommand{\MtWA}{M_{\mathrm{T},\Plp\nu\gamma}}

\def\Li{\mathop{\mathrm{Li}_2}\nolimits}

\newcommand{\Pdbar}{{\bar{\Pd}}}
\newcommand{\Pubar}{{\bar{\Pu}}}

\newcommand{\phz}{\phantom{(0)}}

\def\draftdate{\relax}
\def\mda{\relax}
\def\mua{\relax}
\def\mla{\relax}
\def\draft{
\def\thtystars{******************************}
\def\sixtystars{\thtystars\thtystars}
\typeout{}
\typeout{\sixtystars**}
\typeout{* Draft mode!
         For final version remove \protect\draft\space in source file *}
\typeout{\sixtystars**}
\typeout{}
\def\draftdate{\today}
\def\mua{\marginpar[\boldmath\hfil$\uparrow$]%
                   {\boldmath$\uparrow$\hfil}%
                    \typeout{marginpar: $\uparrow$}\ignorespaces}
\def\mda{\marginpar[\boldmath\hfil$\downarrow$]%
                   {\boldmath$\downarrow$\hfil}%
                    \typeout{marginpar: $\downarrow$}\ignorespaces}
\def\mla{\marginpar[\boldmath\hfil$\rightarrow$]%
                   {\boldmath$\leftarrow $\hfil}%
                    \typeout{marginpar: $\leftrightarrow$}\ignorespaces}
\def\Mua{\marginpar[\boldmath\hfil$\Uparrow$]%
                   {\boldmath$\Uparrow$\hfil}%
                    \typeout{marginpar: $\uparrow$}\ignorespaces}
\def\Mda{\marginpar[\boldmath\hfil$\Downarrow$]%
                   {\boldmath$\Downarrow$\hfil}%
                    \typeout{marginpar: $\downarrow$}\ignorespaces}
\def\Mla{\marginpar[\boldmath\hfil$\Rightarrow$]%
                   {\boldmath$\Leftarrow $\hfil}%
                    \typeout{marginpar: $\leftrightarrow$}\ignorespaces}
\overfullrule 5pt
\oddsidemargin -15mm
\marginparwidth 29mm
}

\titleformat*{\subsubsection}{\large\it}

\numberwithin{equation}{section}

\begin{document}

\thispagestyle{empty}
\def\thefootnote{\fnsymbol{footnote}}
\setcounter{footnote}{1}
\null
\draftdate\hfill FR-PHENO-2014-014

\vfill
\begin{center}
  {\Large \boldmath{\bf NLO QCD and electroweak corrections to $\PW+\gamma$
      production \\[.5em] with leptonic W-boson decays}
\par} \vskip 2.5em
{\large
{\sc Ansgar Denner$^{1}$, Stefan Dittmaier$^{2}$, 
     Markus Hecht$^{2}$, Christian Pasold$^{1}$
}\\[2ex]
{\normalsize \it 
$^1$Julius-Maximilians-Universit\"at W\"urzburg, 
Institut f\"ur Theoretische Physik und Astrophysik, \\
D-97074 W\"urzburg, Germany
}\\[2ex]
{\normalsize \it 
$^2$Albert-Ludwigs-Universit\"at Freiburg, Physikalisches Institut, \\
D-79104 Freiburg, Germany
}\\[2ex]
}
\par \vskip 1em
\end{center}\par
\vskip .0cm \vfill {\bf Abstract:} 
\par 
We present a calculation of the next-to-leading-order electroweak
corrections to $\PW + \gamma$ production, including the leptonic decay
of the W boson and taking into account all off-shell effects of the W~boson, 
where the finite width of the W~boson is implemented using the complex-mass scheme. 
Corrections induced by incoming photons are fully included and find particular
emphasis in the discussion of phenomenological predictions for the LHC.
The corresponding next-to-leading-order QCD corrections are reproduced as well.
In order to separate hard photons from jets, a quark-to-photon fragmentation 
function \'a la Glover and Morgan is employed.
Our results are implemented into Monte Carlo programs allowing for the
evaluation of arbitrary differential cross sections. We present
integrated cross sections for the LHC at $7\TeV$, $8\TeV$, and
$14\TeV$ as well as differential distributions at $14\TeV$ for bare
muons and dressed leptons. Finally, we discuss the impact of anomalous
$WW\gamma$ couplings.

\par
\vskip 1cm
\noindent
December 2014
\par
\null
\setcounter{page}{0}
\clearpage
\def\thefootnote{\arabic{footnote}}
\setcounter{footnote}{0}

\section{Introduction}
\label{se:intro}

Lacking evidence for new physics at the LHC so far, 
precise investigations of
Standard Model (SM) processes are more important than ever. While
scrutinising the properties of the Higgs boson is the most pressing
task, a continued detailed investigation of the electroweak (EW) gauge
bosons has to be pursued as well. Besides single gauge-boson
production in Drell--Yan processes, gauge-boson-pair production offers
various ways of testing the gauge-boson sector of the SM thoroughly.
One of the simplest gauge-boson-pair production processes is the
production of a W boson in association with a photon,
\begin{equation}
\Pp\Pp\to\PW+\ga +X \to \Pl^+ \Pnl + \gamma + X\,.
\end{equation}
Including the branching ratio for the leptonic W~decay, it has a cross
section in the picobarn range and allows for direct tests of the
photon coupling to W bosons. 
Moreover, it constitutes a primary
background to new-physics searches.
Like other di-boson production processes, $\PW\ga$ production has
already been measured at the Tevatron
\cite{Neubauer:2011zz,Abazov:2011rk} and at the LHC
\cite{Chatrchyan:2013fya,Aad:2013izg,Aad:2014fha,Schott:2014awa},
with experimental accuracies of roughly $12\%$
on integrated cross sections.
Up to now, no significant discrepancy between SM predictions
and measurements is seen, a statement that translates into constraints on
anomalous $\PW\PW\gamma$ couplings at the
level of $\sim0.3$ and $\sim 0.04$
for the coupling parameters $\Delta\kappa_\gamma$ and $\lambda_\gamma$,
respectively.
Since the sensitivity to anomalous couplings increases with the
reach in the high-energy tails of distributions, even tighter constraints
are expected from run~2 of the LHC close to its design energy 
of $14\TeV$ with higher luminosity.
Especially the increasing precision of upcoming LHC results asks for
higher precision in predictions.

The leading-order (LO) cross section for $\PW\ga$ production
with on-shell (stable) W~bosons
was published 35 years ago \cite{Brown:1979ux}. The corresponding
next-to-leading-order (NLO) QCD corrections for on-shell $\PW$ bosons
were calculated in \citeres{Smith:1989xz,Ohnemus:1992jn} and
extended to include leptonic decays in the narrow-width approximation
and anomalous couplings in \citere{Baur:1993ir}. Based on full NLO QCD
amplitudes including leptonic decays \cite{Dixon:1998py}, a Monte Carlo
program was presented for $\PW\ga$ production in
\citere{DeFlorian:2000sg}, treating the leptonic decays of the W~boson
in the narrow-width approximation, but retaining all spin information
via decay-angle correlations.  
In this approximation NLO QCD predictions are, for instance, available in the public
program MCFM~\cite{Campbell:2011bn}.
The QCD corrections enhance the cross
section for $\PW\ga$ production at the LHC considerably and thus have
a large impact on the measurement of the $\PW\PW\ga$ coupling.
While gluon-induced NNLO corrections to $\PW\ga$ production including
anomalous couplings were
calculated in \citere{Adamson:2002jb},
recently first results of a complete NNLO calculation have been
published \cite{Grazzini:2014pqa}.
NLO QCD corrections have been interfaced to QCD+QED parton showers
using the POWHEG+MiNLO method \cite{Barze:2014zba}.

It is known since many years that EW corrections can have a sizeable
impact at high energies owing to the presence of logarithmically
enhanced contributions
\cite{Beenakker:1993tt,Beccaria:1998qe,Ciafaloni:1998xg,Kuhn:1999de,Denner:2000jv}.
In particular, distributions of energy-dependent observables may be
affected at the level of several $10\%$.  
The logarithmically
enhanced EW corrections for $\PW\gamma$ production were discussed
in \citere{Accomando:2001fn} and shown to be negative at the level
of $5$--$20\%$. The full EW corrections to $\PW\gamma$ production
including W-boson decays in the pole approximation were presented in
\citere{Accomando:2005ra}.

In this paper we extend the existing calculations for $\PW\gamma$
production in several respects arriving at a complete NLO QCD+EW
calculation for $\Pp\Pp\to\Plp\Pnl\ga+X$. 
To this end, we include the
complete EW one-loop corrections to the partonic processes
$q_i\bar{q}_j\to\Plp\Pnl\gamma$, i.e. we take all off-shell effects of
the W boson into account. Moreover, we include the photon-induced
partonic processes $q_i\ga\to\Plp\Pnl\gamma q_j$,
$\bar{q}_i\ga\to\Plp\Pnl\gamma\bar{q_j}$. 
Like in the calculation of QCD corrections, this requires to separate
hard photons from jets. In order to define this separation in an
infrared-safe way, we use a quark-to-photon fragmentation function
 \cite{Glover:1993qtp,Glover:1994th}.

 This paper is organized as follows: In \refse{se:details} we
 give the details of the calculation, including the general setup and
 the methods used for the virtual and real corrections. Our numerical
 results for total cross sections and distributions in various setups
 as well as with anomalous couplings are presented in
 \refse{se:numres}, while \refse{se:concl} contains our conclusions.

\section{Details of the calculation}
\label{se:details} 

\subsection{General setup}
\label{se:setup} 

The production of a leptonically decaying $\PWp$ boson in combination
with a photon is ruled by quark--antiquark annihilation at LO,
\begin{equation}
\Pu_i \, \Pdbar_j \to \Plp \Pnl \, \gamma \, ,
\label{eq:bornproc}
\end{equation}
where $\Pu_i$ and $\Pdbar_j$ indicate the up-type quarks 
and the down-type antiquarks of the first two generations ($i,j=1,2$). 
The charged lepton and the corresponding neutrino are 
denoted by $l$ and $\Pnl$, 
where $l = \Pe,\mu$.
The LO Feynman diagrams for process \refeq{eq:bornproc} 
are shown in \reffi{fi:bornproc}. 
\begin{figure}[b]
  \begin{center}
\tabcolsep 5pt
    \begin{tabular}{cccc} 
      \input{plots/feynman_diagrams/born/born01.tex} &
      \input{plots/feynman_diagrams/born/born03.tex} &
      \input{plots/feynman_diagrams/born/born04.tex} &
      \input{plots/feynman_diagrams/born/born02.tex} 
    \end{tabular}
  \end{center}
\vspace*{-2em}
  \caption{LO Feynman diagrams for the partonic process $\Pu_i \,\Pdbar_j \to \Plp \Pnl \, \gamma$.}
  \label{fi:bornproc}
\end{figure}
In this work we present NLO corrections to $\PW^+ + \gamma$ production
which can be divided in EW and QCD corrections of the order
$\mathcal{O}(\alpha)$ and $\mathcal{O}(\alpha_s)$, respectively.

We denote the LO cross section calculated with LO parton distribution
functions (PDFs) by $\si^{\mr{LO}}$. The NLO-QCD-corrected cross section is
obtained as 
\begin{align}
  \sigma^{\NLO\,\QCD}&= \sigma^{0} + \Delta\sigma^{\NLO\,\QCD} , \nonumber\\
  \Delta\sigma^{\NLO\,\QCD}&= \sigma^{\alphas}_{\mr{real}}+\sigma^{\alphas}_{\mr{virt}}
                        +\sigma^{\alphas}_{\mr{col}}+\sigma^{\alphas}_{\mr{frag}} ,
\label{eq:sigmaqcd}
\end{align}
where all contributions, including the LO cross section $\sigma^{0}$,
are calculated with NLO PDFs.  The real and the virtual corrections
are given by $\sigma^{\alphas}_{\mr{real}}$ and
$\sigma^{\alphas}_{\mr{virt}}$, respectively, the contribution
$\sigma^{\alphas}_{\mr{col}}$ originates from the redefinition of
the PDFs, and $\sigma^{\alphas}_{\mr{frag}}$ represents the
contribution from fragmentation of a quark into a photon.  All
individual parts in the NLO QCD
contribution $\Delta\sigma^{\NLO\,\QCD}$ are
infrared (IR) divergent and only their sum is IR finite.  While the
separation between $\PW + \gamma$ and $\PW + \mr{jet}$ production is
evident at LO, the final states $\Plp \Pnl \, \gamma \, \Pg$, $\Plp
\Pnl \, \gamma \, \Pd_j$, and $\Plp \Pnl \, \gamma \, \Pubar_i$
appearing in the real NLO corrections to both $\PW + \gamma$ and $\PW
+ \mr{jet}$ production require special care. The technical details of
this aspect are discussed in \refse{suse:fragmentation}.

Analogously to the QCD corrections, 
the EW corrections are given by
\begin{align}
  \Delta\sigma^{\NLO\,\EW}_{\Pq\overline\Pq}&= \sigma^{\alpha}_{\Pq\overline\Pq, \mr{real}}
                                      +\sigma^{\alpha}_{\Pq\overline\Pq, \mr{virt}}
                        +\sigma^{\alpha}_{\Pq\overline\Pq, \mr{col}} ,\nonumber\\
  \Delta\sigma^{\NLO\,\EW}_{\Pq\gamma}&= \sigma^{\alpha}_{\Pq\gamma, \mr{real}}
                        +\sigma^{\alpha}_{\Pq\gamma, \mr{col}}+\sigma^{\alpha}_{\Pq\gamma, \mr{frag}} ,
\end{align}
where the quark--antiquark-induced EW corrections
$\Delta\sigma^{\NLO\,\EW}_{\Pq\overline\Pq}$ and the photon-induced
corrections $\Delta\sigma^{\NLO\,\EW}_{\Pq\gamma}$ are finite, while
their individual contributions are IR divergent.  Unlike the
quark--antiquark- and the quark--gluon-induced channels in the QCD
corrections, $\Delta\sigma^{\NLO\,\EW}_{\Pq\overline\Pq}$ and
$\Delta\sigma^{\NLO\,\EW}_{\Pq\gamma}$ can (in principle)
be distinguished by their
final states.  Analogously to the QCD case, $\sigma^{\alpha}_{ij,
  \mr{real}}$ and $\sigma^{\alpha}_{\Pq\overline\Pq, \mr{virt}}$
denote the real and the virtual corrections, respectively, and
$ij=\Pq\overline\Pq,\Pq\gamma$.  Terms originating from the PDF
redefinition furnish $\sigma^{\alpha}_{ij, \mr{col}}$, and the
fragmentation contribution is described by
$\sigma^{\alpha}_{\Pq\gamma, \mr{frag}}$.
Note that no fragmentation contribution is required in the $\Pq\overline\Pq$~channel
at NLO EW, because there is no jet in the final state in this order.

We choose to combine QCD and EW corrections using the naive product
\begin{align}\label{naive-product}
  \sigma^{\mr{NLO}}&= \sigma^{\mr{LO}}\klam{1+\delta_{\QCD}}
                                  \klam{1+\delta_{\EW,\Pq\overline{\Pq}}+\delta_{\EW, \Pq\gamma}}
\nonumber
\\
  &= \sigma^{\mr{NLO\,QCD}} \klam{1+\delta_{\EW,\Pq\overline{\Pq}}+\delta_{\EW, \Pq\gamma}},
\end{align}
where the relative QCD, EW, and photon-induced corrections are defined by
\begin{align}
 \de_{\QCD} & = \frac{\si^{\NLO\,\QCD}-\si^{\mr{LO}}}{\si^{\mr{LO}}} , &
 \de_{\EW, \Pq\overline{\Pq}} & = \frac{\Delta\sigma^{\NLO\,\EW}_{\Pq\overline\Pq}}{\si^{0}} , &
 \de_{\EW, \Pq\gamma} & = \frac{\Delta\sigma^{\NLO\,\EW}_{\Pq\gamma}}{\si^{0}} ,
\label{eq:relcor}
\end{align}
respectively. While the relative QCD corrections are normalized to the
LO cross section $\si^{\mr{LO}}$, calculated with LO PDFs,
the EW corrections are normalized to the LO cross section $\si^{0}$,
calculated with NLO PDFs. By this definition,
$K_{\QCD}=1+\delta_{\QCD}$ is the standard QCD factor,
and the relative EW corrections are practically independent of the PDF
set.

In order to calculate the necessary amplitudes in the 
't~Hooft--Feynman gauge we use traditional methods based on Feynman
diagrams. Two independent calculations were performed to guarantee the
accuracy of the presented results. In both cases we employ the
Weyl--van-der-Waerden spinor formalism as formulated in
\citere{Dittmaier:1998nn} for the numerical evaluation of the
amplitudes. For the numerical calculation of the loop integrals
the \texttt{COLLIER} library \cite{Denner:2014gla} is used, which is
based on the results of
\citeres{Denner:2002ii,Denner:2005nn,Denner:2010tr} and involves two
different independent implementations of all one-loop integrals.

One amplitude calculation is based on the program \texttt{POLE}
\cite{Accomando:2001fn}, which internally uses \texttt{FEYNARTS}~3
\cite{Hahn:2000kx,Hahn:2001rv} and \texttt{FORMCALC}
\cite{Hahn:1998yk} for the generation of the amplitudes. The numerical
integration is performed by the multi-channel phase-space generator
\texttt{LUSIFER} \cite{Dittmaier:2002ap}  extended to use
\texttt{Vegas}~\cite{Lepage:1977sw,Lepage:1980dq} in order to optimize
each phase-space mapping.

In the second calculation the virtual amplitudes are generated by 
\texttt{FEYNARTS}~1~\cite{Kublbeck:1990xc} and algebraically reduced
with an in-house \texttt{MATHEMATICA} package,
automatically transferring the results into a \texttt{FORTRAN} 
code.
For the numerical evaluation of integrated and differential cross
sections the amplitudes are implemented into a \texttt{FORTRAN}
program using the \texttt{VEGAS} algorithm for a proper numerical
integration.  In case of $\PW + \gamma$ production sharp
resonances appear, 
demanding additional phase-space mappings.
Therefore, analytical Breit--Wigner mappings are introduced in the 
phase-space parametrization, allowing for a stable
numerical integration by flattening the integrand.

\subsection{Virtual corrections}
\label{se:virt} 

In this section we shortly discuss the calculation of the virtual QCD
and EW corrections to the partonic process \refeq{eq:bornproc}. The
QCD corrections receive contributions from self-energy, vertex, and box
(4-point) diagrams only.  The NLO EW corrections in addition involve 
pentagon (5-point) diagrams.  The structural diagrams for the NLO EW
corrections are shown in \reffis{fi:selfenergies}--\ref{fi:pentagons}. 
There are about 280 EW one-loop diagrams involving 50 box and 16
pentagon diagrams (presented in \reffi{fi:explicitpentagons}).%
\begin{figure}
  \begin{center}
\tabcolsep 5pt
    \begin{tabular}{ccccc} 
      \input{plots/feynman_diagrams/virt/self02.tex} &
      \input{plots/feynman_diagrams/virt/self01.tex} &
      \input{plots/feynman_diagrams/virt/self04.tex} &
      \input{plots/feynman_diagrams/virt/self03.tex} &
      \input{plots/feynman_diagrams/virt/self07.tex} 
      \\
      \input{plots/feynman_diagrams/virt/self08.tex} &
      \input{plots/feynman_diagrams/virt/self09.tex} &
      \input{plots/feynman_diagrams/virt/self10.tex} &
      \input{plots/feynman_diagrams/virt/self05.tex} &
      \input{plots/feynman_diagrams/virt/self06.tex} 
    \end{tabular}
  \end{center}
\vspace*{-2em}
  \caption{Self-energy corrections to the partonic process $\Pu_i \,\Pdbar_j \to \Plp \Pnl \, \gamma$.}
  \label{fi:selfenergies}
%
\vspace*{1em}
  \begin{center}
\tabcolsep 5pt
    \begin{tabular}{ccccc}
      \input{plots/feynman_diagrams/virt/tri13.tex} &
      \input{plots/feynman_diagrams/virt/tri15.tex} &
      \input{plots/feynman_diagrams/virt/tri01.tex} &
      \input{plots/feynman_diagrams/virt/tri06.tex} &
      \input{plots/feynman_diagrams/virt/tri05.tex} 
      \\
      \input{plots/feynman_diagrams/virt/tri12.tex} &
      \input{plots/feynman_diagrams/virt/tri09.tex} &
      \input{plots/feynman_diagrams/virt/tri07.tex} &
      \input{plots/feynman_diagrams/virt/tri10.tex} &
      \input{plots/feynman_diagrams/virt/tri11.tex} 
      \\
      \input{plots/feynman_diagrams/virt/tri08.tex} &
      \input{plots/feynman_diagrams/virt/tri14.tex} &
      \input{plots/feynman_diagrams/virt/tri02.tex} &
      \input{plots/feynman_diagrams/virt/tri04.tex} &
      \input{plots/feynman_diagrams/virt/tri03.tex} 
    \end{tabular}
  \end{center}
\vspace*{-2em}
  \caption{Vertex corrections to the partonic process $\Pu_i \,\Pdbar_j \to \Plp \Pnl \, \gamma$.}
  \label{fi:triangles}
%
\vspace*{1em}
  \begin{center}
\tabcolsep 5pt
    \begin{tabular}{ccccc}
      \input{plots/feynman_diagrams/virt/box01.tex} &
      \input{plots/feynman_diagrams/virt/box04.tex} &
      \input{plots/feynman_diagrams/virt/box02.tex} &
      \input{plots/feynman_diagrams/virt/box03.tex} &
      \input{plots/feynman_diagrams/virt/box05.tex}
    \end{tabular}
  \end{center}
\vspace*{-2em}
  \caption{Box corrections to the partonic process $\Pu_i \,\Pdbar_j \to \Plp \Pnl \, \gamma$.}
  \label{fi:boxes}
\end{figure}%
\begin{figure}
  \begin{center}
\tabcolsep 5pt
    \begin{tabular}{c}
      \input{plots/feynman_diagrams/virt/pent01.tex}
    \end{tabular}
  \end{center}
\vspace*{-2em}
  \caption{Pentagon corrections for the partonic process $\Pu_i \,\Pdbar_j \to \Plp \Pnl \, \gamma$.}
  \label{fi:pentagons}
%
\vspace*{1em}
  \begin{center}
\tabcolsep 5pt
    \begin{tabular}{ccccc}
      \input{plots/feynman_diagrams/virt/exppent02.tex} &
      \input{plots/feynman_diagrams/virt/exppent04.tex} &
      \input{plots/feynman_diagrams/virt/exppent05.tex} &
      \input{plots/feynman_diagrams/virt/exppent06.tex} &
      \input{plots/feynman_diagrams/virt/exppent10.tex}
      \\
      \input{plots/feynman_diagrams/virt/exppent07.tex} &
      \input{plots/feynman_diagrams/virt/exppent09.tex} &
      \input{plots/feynman_diagrams/virt/exppent08.tex} &
      \input{plots/feynman_diagrams/virt/exppent01.tex} &
      \input{plots/feynman_diagrams/virt/exppent03.tex} 
    \end{tabular}
  \end{center}
\vspace*{-2em}
  \caption{Explicit pentagon diagrams for the partonic process $\Pu_i \,\Pdbar_j \to \Plp \Pnl \, \gamma$.}
  \label{fi:explicitpentagons}
\end{figure}%

For the proper description of the resonant W-boson propagators we use
the complex-mass scheme
\cite{Denner:1999gp,Denner:2005fg,Denner:2006ic}. In this scheme the
W- and Z-boson masses are consistently treated as complex quantities,
defined as the locations of the propagator poles in the complex plane.
As a consequence the electroweak mixing angle and many couplings
become complex as well. The complex-mass scheme fully respects gauge
invariance, i.e.\ the underlying Ward identities are fulfilled and
no dependence on the gauge-fixing procedure remains.

The one-loop amplitude can be represented in terms of standard matrix
elements, which comprise all polarization-dependent quantities, and
invariant coefficients, which are linear combinations of
tensor-integral coefficients \cite{Denner:1991kt}. The tensor integral
coefficients are evaluated with the library \texttt{COLLIER}
\cite{Denner:2014gla}. The tensor integrals are recursively reduced to
master integrals 
at the numerical level. 
All 5-point functions are directly reduced to 4-point functions based
on the methods of
\citeres{Denner:2002ii,Denner:2005nn,Melrose:1965kb}.  The reduction
to 4-point integrals uses the Passarino--Veltman algorithm
\cite{Passarino:1978jh} as well as dedicated expansion methods for
exceptional phase-space regions \cite{Denner:2005nn}.
The one-loop scalar integrals are evaluated with 
complex masses using the methods and results of
\citeres{Denner:2010tr,'tHooft:1978xw,Denner:1991qq,Beenakker:1988jr}.

Ultraviolet divergences are regularized dimensionally. 
For IR singularities (soft and/or collinear) we either use pure dimensional regularization
or alternatively pure mass regularization with infinitesimal gluon and
photon masses and small fermion masses, which are only kept in the
arguments of mass-singular logarithms. The results obtained with the two
different IR regularizations are in perfect numerical agreement.

We use the on-shell renormalization as
described in \citere{Denner:2005fg} for the complex-mass scheme. 
The strong coupling constant is renormalized in the
$\overline{\mathrm{MS}}$ scheme with five active flavours 
with the top quark
decoupled from the running of the strong coupling constant.

\subsection{Real corrections}
\label{se:real} 

The real corrections are induced by the
radiation of an additional photon or 
QCD parton.  In order to isolate the
soft and collinear divergences in 
the phase-space integration, the
dipole subtraction formalism is applied. Specifically we use the QCD
dipoles introduced in 
\citeres{Catani:1996vz,Catani:2002hc} for the calculation of
the real QCD corrections and the QED dipoles introduced in
\citeres{Dittmaier:EWsubtraction,Dittmaier:2008md} for the evaluation of the real EW
corrections. First we focus on the real EW corrections, where we
distinguish between collinear-safe
and non-collinear-safe observables. 
Then we discuss the quark--photon-induced channels and
the real QCD corrections.  As already mentioned, at NLO a simple
separation of $\PW+\gamma$ and $\PW+\mr{jet}$ production is not
possible. Therefore, the concept of democratic clustering and the
quark-to-photon fragmentation function are introduced, allowing for a
well-defined separation of the two processes.

\subsubsection{Real EW corrections}

\begin{figure}
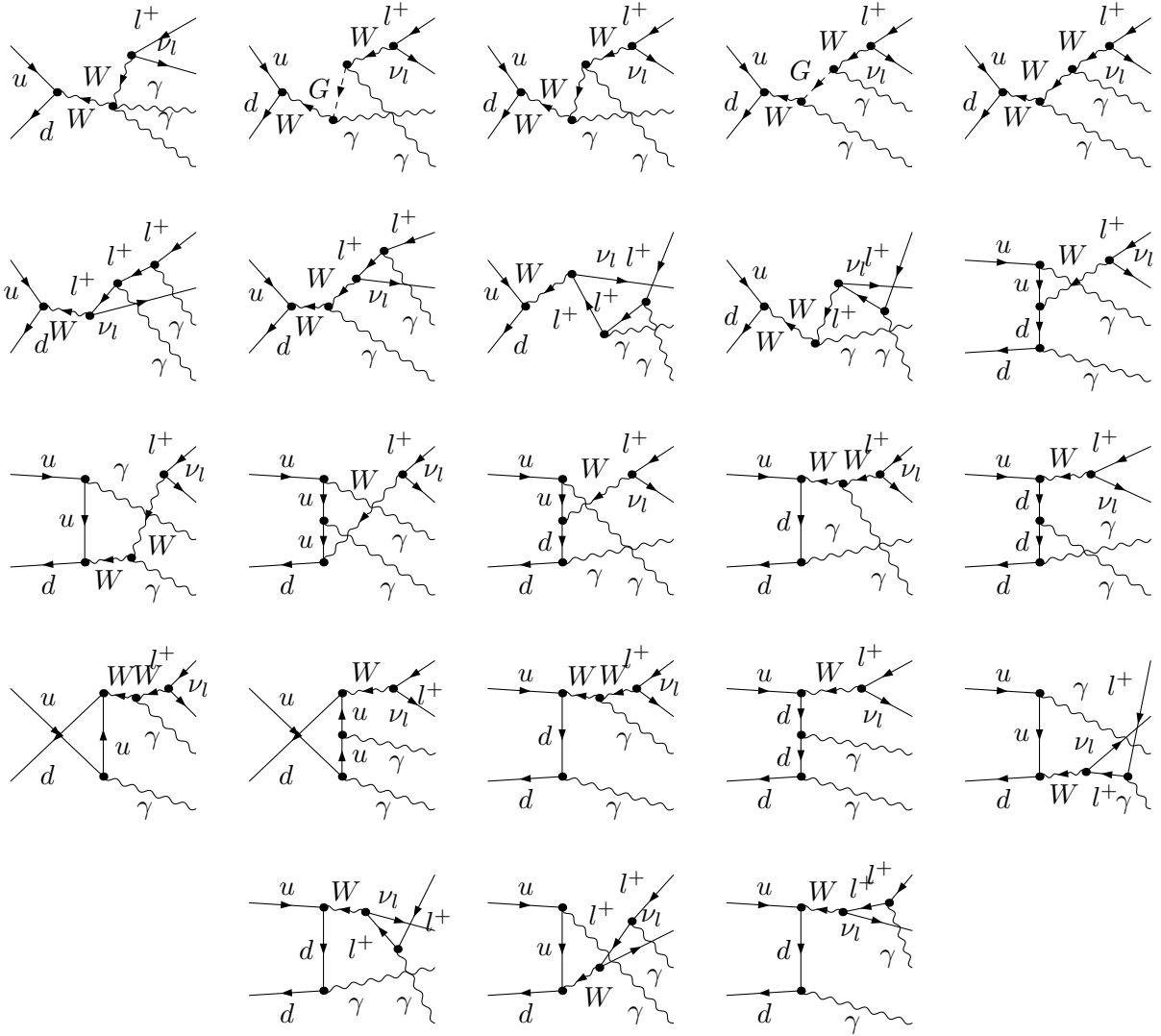
     
  \begin{center}
\tabcolsep 5pt
    \begin{tabular}{ccccc}
      \input{plots/feynman_diagrams/real_qqEW/qqEW_real01.tex} &
      \input{plots/feynman_diagrams/real_qqEW/qqEW_real02.tex} &
      \input{plots/feynman_diagrams/real_qqEW/qqEW_real03.tex} &
      \input{plots/feynman_diagrams/real_qqEW/qqEW_real04.tex} &
      \input{plots/feynman_diagrams/real_qqEW/qqEW_real05.tex} 
      \\
      \input{plots/feynman_diagrams/real_qqEW/qqEW_real06.tex} &
      \input{plots/feynman_diagrams/real_qqEW/qqEW_real07.tex} &
      \input{plots/feynman_diagrams/real_qqEW/qqEW_real08.tex} &
      \input{plots/feynman_diagrams/real_qqEW/qqEW_real09.tex} &
      \input{plots/feynman_diagrams/real_qqEW/qqEW_real10.tex}
      \\
      \input{plots/feynman_diagrams/real_qqEW/qqEW_real11.tex} &
      \input{plots/feynman_diagrams/real_qqEW/qqEW_real12.tex} &
      \input{plots/feynman_diagrams/real_qqEW/qqEW_real13.tex} &
      \input{plots/feynman_diagrams/real_qqEW/qqEW_real14.tex} &
      \input{plots/feynman_diagrams/real_qqEW/qqEW_real15.tex}
      \\
      \input{plots/feynman_diagrams/real_qqEW/qqEW_real16.tex} &
      \input{plots/feynman_diagrams/real_qqEW/qqEW_real17.tex} &
      \input{plots/feynman_diagrams/real_qqEW/qqEW_real18.tex} &
      \input{plots/feynman_diagrams/real_qqEW/qqEW_real19.tex} &
      \input{plots/feynman_diagrams/real_qqEW/qqEW_real20.tex}
      \\
      & 
      \input{plots/feynman_diagrams/real_qqEW/qqEW_real21.tex} &
      \input{plots/feynman_diagrams/real_qqEW/qqEW_real22.tex} &
      \input{plots/feynman_diagrams/real_qqEW/qqEW_real23.tex} &
    \end{tabular}
    
  \end{center}
\vspace*{-2em}
  \caption{Feynman diagrams of the quark--antiquark-induced real EW corrections for the partonic process $\Pu_i \,\Pdbar_j \to \Plp \Pnl \, \gamma \, \gamma$.
           }
  \label{fi:qqEW}
\end{figure}

We first focus on the quark--antiquark-induced EW corrections where  
the final state contains two photons,
\begin{equation}
\Pu_i \,\Pdbar_j \to \Plp \Pnl \, \gamma \, \gamma \, .
\label{eq:bremsprocEW1}
\end{equation}
The corresponding Feynman diagrams are shown in \reffi{fi:qqEW}.
These contributions include soft and collinear singularities.
While the soft singularities completely cancel against corresponding 
contributions in the virtual corrections, the cancellation of collinear
singularities between real and virtual corrections is only partial in general.
In the following we separately discuss the treatment of collinear
singularities originating from photon radiation off initial-state (IS)
or final-state (FS) 
leptons.

Following the standard treatment in the QCD-improved parton model,
singularities connected to collinear splittings of IS partons
are considered to be part of the incoming proton and are
absorbed into the PDFs ($f_{\Pq}$) by the
redefinition (see, e.g., \citeres{Baur:1998kt,Dittmaier:2009cr})
\begin{align}
  f_{\Pq}\klam{x} 
\rightarrow\; &  f_{\Pq}\klam{x,\mu^{2}_{\mr{F}}} 
          - \frac{\alpha}{2\pi}Q^{2}_{\Pq}
\int^{1}_{x} \frac{\md z}{z}\, f_{\Pq}\klam{\frac{x}{z},\mu^{2}_{\mr{F}}}
\nonumber\\
          & \times\left\{\ln{\klam{\frac{\mu^{2}_{\mr{F}}}{m^{2}_{q}}}} \left[P_{ff}\klam{z}\right]_{+}
          -\left[P_{ff}\klam{z} \klam{2\ln{\klam{1-z}}+1}\right]_{+}+C_{ff}\klam{z}\right\},
\end{align}
where $x$ is the energy fraction carried by the parton coming
from a proton, $\mu_{\mr{F}}$ is the factorization scale, and $m_{\Pq}$
and $Q_{\Pq}$ are the quark mass and charge, respectively. The
splitting function is given by
\begin{align}\label{eq:AP-function}
  P_{ff}\klam{z}&=\frac{1+z^{2}}{1-z},
\end{align}
where $[\ldots]_+$ denotes the usual $(+)$-distribution prescription
\begin{align}
\int_0^1\rd x\, \Big[f(x)\Big]_+ g(x) =
\int_0^1\rd x\, f(x) \left[g(x)-g(1)\right],
\end{align}
with $g(x)$ representing a smooth test function.

The coefficient function $C_{ff}$ defines the factorization scheme.
Actually the ${\cal O}(\alpha)$-corrected NLO PDF set NNPDF23~\cite{Ball:2013hta} 
is only of LO with respect to QED corrections, i.e.\ they do not uniquely define
a factorization scheme, but they should be most
adequately used in a DIS-like factorization scheme for QED corrections
(see \citere{Diener:2005me} for arguments), so that
the coefficient function reads
\begin{align}
  C^{\mr{DIS}}_{ff}\klam{z} &
     = \left[P_{ff}\klam{z}\klam{\ln\klam{\frac{1-z}{z}}-\frac{3}{4}}+\frac{9+5z}{4}\right]_{+}.
\end{align}

The cancellation 
of collinear singularities connected with
photon emission of FS
leptons depends on the 
level of inclusiveness in the event reconstruction.
Trying to simulate the experimental setup as close as possible, two
scenarios are considered.
The first
scenario is typical for a FS electron, which experimentally cannot be
separated from a collinear photon. In the electromagnetic calorimeter
only an electromagnetic shower is detected that is declared as a
charged lepton. Technically this means that the charged lepton and the
photon are recombined to one particle in the collinear regime, leading
to a complete cancellation of the corresponding 
IR singularities, as dictated by the KLN
theorem \cite{Kinoshita:1962ur,Lee:1964is}. Therefore this case is
called the collinear-safe (CS) case.  The second scenario applies to a
FS muon, which is detected in the muon chamber, while a collinear
photon is absorbed in the electromagnetic calorimeter. Thus, in the
calculation we should not recombine the muon and the photon even if
they are produced collinearly.  
This separation of the photon and the ``bare'' muon has an important
implication on the energy fraction 
\begin{align}
  z_{f}=\frac{p^{0}_{f}}{p^{0}_{f}+k^{0}_{\gamma}},
\label{eq:zf}
\end{align}
of a charged fermion $f$ produced together with a photon in a fixed
collinear cone, 
where $p^{0}_{f}$ and $k^{0}_{\gamma}$ are the energies of the charged
lepton and the photon, respectively. 
The absence of photon recombination
in general disturbs the inclusive integration over $z_f$ in observables,
which, however, is required for the cancellation of collinear
singularities between the real and the virtual corrections.
In this case
mass-singular contributions of the form $\alpha\ln\klam{m_{f}}$
remain, and this scenario is called the non-collinear-safe (NCS) case.

In contrast to the virtual corrections where the IR singularities
appear as analytic expressions after performing the loop integrations, 
in the real corrections the IR singularities emerge 
from the phase-space integration of the FS particles,
which is best done numerically. In order to
analytically extract those singularities from the numerical integration 
to allow for an analytical cancellation,
the dipole subtraction method is used. This technique was introduced in
\citere{Dittmaier:EWsubtraction} for CS photon radiation and extended to the
NCS case in \citere{Dittmaier:2008md}. We only give a short overview of the latter
case, which covers the CS case as well.

To extract the IR singularities from the real corrections, a so-called
subtraction function $\abs{\M_{\sub}}^{2}$ is subtracted from the
squared real-emission amplitude cancelling all IR singularities and
allowing for a stable numerical integration. The subtraction function
is constructed out of the squared Born-level matrix element
$\abs{\M_{0}}^{2}$ and the dipole functions
$g^{\klam{\mr{sub}}}_{ff'}$,
\begin{align}
\label{subtractionfunction}
  \abs{\M_{\sub}(\Phi_{1})}^{2}&= -\sum_{f \neq f'}Q_{f}\sigma_{f}Q_{f'}\sigma_{f'}e^{2}
                              g^{\klam{\mr{sub}}}_{ff'}\klam{p_{f},p_{f'},k}
                               \abs{\M_{0}\klam{{\tilde{\Phi}_{0,ff'}}}}^{2},
\end{align}
where $p_{f}$, $p_{f'}$, and $k$ are the emitter, the spectator, and
the photon momenta, respectively. The projection of the
$\klam{N+1}$-particle phase space $\Phi_{1}$ to the $N$-particle phase
space is denoted as $\tilde{\Phi}_{0,ff'}$.  The charges of the
emitter and the spectator are given by $Q_{f}$ and $Q_{f'}$,
respectively, and the sign factors $\sigma_{f/f'}$ indicate the charge
flows taking
the values $+1(-1)$ for incoming (outgoing) fermions
or outgoing (incoming) antifermions. Adding the subtraction function
back and integrating 
it over the one-particle phase space containing the IR singularity, the
IR singularity can be cancelled against the corresponding
singularity from the
virtual corrections analytically.  As already mentioned, mass-singular
contributions remain in the NCS case, since we cut on the energy
fraction $z_{f}$ of the charged lepton inside the photon radiation
cone by histogram binning or by applying event-selection cuts.  In
practice this means that the dependencies on $z_{f}$ have to be made
explicit during the whole subtraction procedure.  For the application
of event-selection cuts, the momenta that enter the LO matrix element
inside the subtraction function have to be transferred to an
$(N+1)$-particle phase space with photon emission by the assignment
\begin{align}
  p_{f}&\to z_{ff'}\tilde{p}^{\klam{ff'}}_{f},& k&\to (1-z_{ff'})\tilde{p}^{\klam{ff'}}_{f},
& p_{f'}&\to \tilde{p}^{\klam{ff'}}_{f'},
\label{eq:dipmomenta}
\end{align}
which splits the momentum $\tilde{p}^{\klam{ff'}}_{f}$ of the emitter fermion $f$
into the momenta $p_{f}$ and $k$ of a collinear pair of a fermion and a photon, respectively.
The energy fraction $z_{ff'}$ defining this splitting tends to $z_f$ of \refeq{eq:zf}
in the corresponding singular limit.  
With this the subtraction proceeds as
follows,
\begin{align}
\int \, \rd \Phi_1 &\biggl[|\M_{\mr{real}}|^2 \Theta_{\cut} 
(p_f,k,p_{f'},\{k_n\}) \nonumber \\
& - \sum_{f \neq f'} |\M_{\sub,f f'}|^2 \Theta_{\cut} \left( z_{f f'} 
\tilde{p}_{f}^{(f f')}, (1 - z_{f f'}) \tilde{p}_{f}^{(f f')}, 
\tilde{p}_{f'}^{(f f')}, \{\tilde{k}_n\} \right) \biggr] \, ,
\label{eq:NCSreal}
\end{align}
where the sum runs over all emitter--spectator pairs $f,f'$. 
The momenta of the remaining particles are denoted by
$\{\tilde{k}_n\}$. The cut function
$\Theta_{\cut}$ acting on the $(N+1)$-particle phase space equals one
if the configuration of final-state momenta passes the cuts, and zero
otherwise. 

Adding the subtraction function 
$
\int \rd \Phi_{1} |\M_{\sub,ff'}(\Phi_{1})|^2
$
back and integrating out the divergent part analytically, the dependence 
on $z_{f}$ has to be kept explicit. Comparing with the CS case this 
leads to additional terms including a $(+)$-distribution acting on $z_f$ and 
in case of an IS spectator a double $(+)$-distribution 
$\left[\dots\right]^{\klam{x,y}}_{+}$ defined by
\begin{align}
\int_0^1\rd x\, \int_0^1\rd y\, 
\Big[f(x,y)\Big]^{\klam{x,y}}_+ g(x,y) =
\int_0^1\rd x\, f(x,y) 
\left[g(x,y)-g(1,y)-g(x,1)+g(1,1)\right],
\end{align}
where $g(x,y)$ is a smooth test function. 
In the following we 
only discuss the case of an IS 
spectator, since 
it is the most complicated. 
More details as well as the treatment of the
remaining emitter--spectator cases can be found in \citere{Dittmaier:2008md}.
The NCS extension of the CS subtraction procedure is constructed in such a way
that the $z_{f}$-dependent contributions are added in a straight-forward
manner. These additional terms, which become zero in the CS case, 
include the mass singularity that remains due to the incomplete 
cancellation of IR divergences originating from the virtual corrections. 
The final equation for the integrated subtraction function with a 
FS emitter and an IS spectator reads
\begin{align}
\int & \rd \Phi_{1} |\M_{\sub,i a}(\Phi_{1})|^2 \, =  \, - 
\frac{\alpha}{2 \pi} Q_{i} \sigma_{i} Q_{a} \sigma_{a} \int^{1}_{0} \rd x \int 
\rd \tilde{\Phi}_{\mathrm{0},ia} (P^2_{ia},x) \int^{1}_{0} \rd z 
\nonumber \\
&\times \Theta_{\cut} \left( p_{i} = z \tilde{p}_{i} (x), k = 
(1-z)\tilde{p}_{i} (x),\{ \tilde{k}_{n} (x)\} \right) \nonumber \\
&\times \frac{1}{x} \Big\{ G^{(\sub)}_{ia} (P^2_{ia}) \, 
\delta (1-x) \, \delta (1-z) + \left[ 
{\cal{G}}^{(\sub)}_{ia} (P^2_{ia},x) \right]_{+} \delta 
(1-z) \Big. \nonumber \\
&\Big. \hspace{1cm} + \left[ \bar{\cal{G}}^{(\sub)}_{ia} 
(P^2_{ia},z) \right]_{+} \delta (1-x) + \left[ 
\bar{g}^{(\sub)}_{ia} (x,z) \right]^{\klam{x,z}}_{+} \Big\}
\Big| \M_{\mathrm{0}} (\tilde{p}_{i} (x),\tilde{p}_{a} (x),
\{\tilde{k}_{n} (x)\}) \Big|^{2} \, ,
\label{eq:NCSreadded}
\end{align}
where the fine-structure constant is defined by $\alpha = e^2 / (4
\pi)$ and the indices $i$ and $a$ indicate the FS emitter and the IS
spectator, respectively. The original momentum of the IS spectator is
re-scaled by $x$ ($\tilde{p}_{i}=x p_{i}$).  Therefore,
$\rd\tilde{\Phi}_{\mathrm{0},ia}(P^2_{ia},x)$, which indicates 
the phase-space
measure of the momenta $\tilde{p}_{i}(x)$, $\tilde{p}_{i} (x)$, and $\{
\tilde{k}_{n} (x)\}$ before FS radiation, is $x$ dependent.
Due to the $(+)$-distribution in $x$, two sets of phase-space momenta,
$\tilde{\Phi}_{\mathrm{0},ia}(P^2_{ia},x)$ and
$\tilde{\Phi}_{\mathrm{0},ia}(P^2_{ia},1)$ have to be generated per
phase-space point. 
The invariant $P^2_{ia} = ({p}_{i} - {p}_{a}+k)^{2} =
(\tilde{p}_{i} - \tilde{p}_{a})^{2}$ has to be evaluated with the
momenta entering the corresponding LO matrix element
$\abs{\M_{\mathrm{0}}}^{2}$. The function $\Theta_{\cut}$ represents
the application of the phase-space cuts on the $(N+1)$-particle phase
space, where the momenta $p_{i}$ and $k$ are reconstructed from the
momentum $\tilde{p}_{i}(x)$ as given in \refeq{eq:dipmomenta}. 
While line~3 of \refeq{eq:NCSreadded} describes the $z$-independent 
contributions which also appear in the CS case, line~4 contains the 
additional $z$-dependent terms emerging in the NCS scenario.
In the case of a FS emitter and an IS
spectator the endpoint function $G^{(\sub)}_{ia} (P^2_{ia})$ and the
distribution ${\cal{G}}^{(\sub)}_{ia} (P^2_{ia},x)$ in the limit of
$m_i \rightarrow 0$ are given by
\begin{align}
G^{(\sub)}_{ia} (P^2_{ia}) &= \L (|P^2_{ia}|,m_i^2) - 
\frac{\pi^2}{2} + \frac{3}{2} \, , \nonumber\\
{\cal{G}}^{(\sub)}_{ia} (P^2_{ia},x) &= \frac{1}{1 -x} 
\left[ 2 \ln \left( \frac{2 - x}{1 -x} - \frac{3}{2} \right) 
\right] \, , 
\label{eq:csfunctions}
\end{align}
with
\begin{equation}
\L (P^2,m^2) = \ln \left( \frac{m^2}{P^2} \right) \ln \left( 
\frac{m_{\gamma}^2}{P^2} \right) + \ln \left( \frac{m_{\gamma}^2}{P^2} 
\right) - \frac{1}{2} \ln^2 \left( \frac{m^2}{P^2} \right) + 
\frac{1}{2}\ln \left( \frac{m^2}{P^2} \right) \, .
\label{eq:Lfunction}
\end{equation}
The functions containing the $z$ dependence read 
\begin{align}
\bar{g}^{(\sub)}_{ia} (x,z) &= \frac{1}{1-x} \left( 
\frac{2}{2-x-z} -1 -z \right) \, , \nonumber\\
\bar{\cal{G}}^{(\sub)}_{ia} (P^2_{ia},z) &= P_{ff} (z) \left[ \ln 
\left( \frac{- P^2_{ia} z}{m_i^2} \right) - 1 \right] - 
\frac{2\ln(2-z)}{1-z} + (1+z)\ln(1-z) + (1 - z) \, ,
\label{eq:fininncsfunctions}
\end{align}
where the Altarelli--Parisi splitting function $P_{ff} (z)$ is defined
in \refeq{eq:AP-function}.
The remaining fermion-mass dependence originating
from the incomplete
cancellation of collinear singularities is included in the function
$\bar{\cal{G}}^{(\sub)}_{ia} (P^2_{ia},z)$.

\subsubsection{Real QCD and photon-induced corrections}

The real QCD and the real photon-induced corrections have the same FS
signature and cause the same problem when a photon and a jet become
collinear. Therefore, we discuss these corrections together. 
At NLO QCD the
following channels contribute to the real corrections,
\begin{align}
\Pu_i \,\Pdbar_j &\to \Plp \Pnl \, \gamma \, \Pg \, ,\nonumber\\
\Pu_i \,\Pg      &\to \Plp \Pnl \, \gamma \, \Pd_j \, ,\nonumber\\
\label{eq:bremsprocQCD3}
\Pdbar_j \, \Pg   &\to \Plp \Pnl \, \gamma \, \Pubar_i.
\end{align}
The corresponding Feynman diagrams for the quark--antiquark-induced
real corrections are shown in \reffi{fi:qqQCD}. 
\begin{figure}     
  \begin{center}
\tabcolsep 5pt
    \begin{tabular}{ccccc}
      \input{plots/feynman_diagrams/real_qqQCD/qqQCD_real05.tex} &
      \input{plots/feynman_diagrams/real_qqQCD/qqQCD_real07.tex} &
      \input{plots/feynman_diagrams/real_qqQCD/qqQCD_real01.tex} &
      \input{plots/feynman_diagrams/real_qqQCD/qqQCD_real02.tex} &
      \input{plots/feynman_diagrams/real_qqQCD/qqQCD_real06.tex} 
      \\
      \input{plots/feynman_diagrams/real_qqQCD/qqQCD_real08.tex} &
      \input{plots/feynman_diagrams/real_qqQCD/qqQCD_real04.tex} &
      \input{plots/feynman_diagrams/real_qqQCD/qqQCD_real03.tex} &
      \input{plots/feynman_diagrams/real_qqQCD/qqQCD_real09.tex} &
      \input{plots/feynman_diagrams/real_qqQCD/qqQCD_real10.tex}
    \end{tabular}
  \end{center}
\vspace*{-2em}
  \caption{Feynman diagrams of the quark--antiquark-induced real QCD corrections for the partonic process $\Pu_i \,\Pdbar_j \to \Plp \Pnl \, \gamma \, \Pg$.
          }
  \label{fi:qqQCD}
\end{figure}
The quark--gluon-induced channels can be obtained from the quark--antiquark-induced
channel by crossing the gluon into the IS and the (anti)quark into the
FS.  For the extraction of the IR singularities from the real emission
amplitudes the Catani--Seymour dipole-subtraction formalism is applied
\cite{Catani:1996vz,Catani:2002hc}.

Calculating the photon-induced corrections the following 
channels have to be taken into account,
\begin{align}
& \Pu_i \,\gamma      \to \Plp \Pnl \, \gamma \, \Pd_j \, ,\nonumber\\
\label{eq:bremsprocEW3}
& \Pdbar_j \, \gamma   \to \Plp \Pnl \, \gamma \, \Pubar_i \, .
\end{align}
Collinear singularities from IS splittings are
compensated by a redefinition of the PDFs 
(see, e.g., \citere{Dittmaier:2009cr}),
\begin{align}
  f_{\Pq}\klam{x} \rightarrow f_{\Pq}\klam{x,\mu^{2}_{\mr{F}}} 
          - \frac{\alpha}{2\pi}3Q^{2}_{\Pq}\int^{1}_{x} \frac{\md z}{z}\, f_{\gamma}\left(\frac{x}{z},\mu^{2}_{\mr{F}}\right)
           \left\{\ln\klam{\frac{\mu^{2}_{\mr{F}}}{m^{2}_{q}}}P_{f\gamma}\klam{z}+
            C_{f\gamma}\right\},
\end{align}
where the splitting function is defined as
\begin{align}
  P_{f\gamma}\klam{z}=z^{2}+(1-z)^{2}, 
\end{align}
and the factorization-scale-dependent coefficient function $C_{f\gamma}(z)$
evaluated in the DIS scheme reads 
\begin{align}
  C^{\mr{DIS}}_{f\gamma}(z)=P_{f\gamma}(z)\ln \klam{\frac{1-z}{z}} - 8z^{2}+8z-1\,.
\end{align}
For extracting the collinear IS singularities from the real emission amplitude,
the dipole subtraction formalism introduced in \citere{Dittmaier:2008md} is applied. 
Demanding a photon in the FS requires that the photon and the FS
jet are separated in a well-defined way. This causes an additional 
complication that is discussed in the next section.

\subsubsection{Quark-to-photon fragmentation function}
\label{suse:fragmentation}

At NLO we have to deal with photons that are radiated collinearly to a
jet.  Experimentally it is not possible to decide whether the photon
comes from the hard scattering process or is generated during the
hadronization of a quark or a gluon. Furthermore, a collinear
divergence appears if the photon is radiated collinear to a FS quark.
This divergence cancels if photons and QCD partons are treated on
equal footing, i.e.\ if photon recombination with QCD partons is
included in the jet algorithm.  However, in such a procedure the two
processes of $\PW+\gamma$ and $\PW+\mr{jet}$ production are not
separated at all, so that valuable experimental information would
remain unexploited.  To separate the two processes, a well-defined
procedure is needed to differentiate between collinear photons and
jets.  Note that simply imposing a separation cut between the photon
and the jet would spoil the cancellation of IR divergences.

The problem of separating the two processes can
be solved by introducing the concept of democratic clustering and a
quark-to-photon fragmentation function as suggested in
\citeres{Glover:1993qtp,Glover:1994th}.  This concept treats photons
and partons on equal footing, meaning that they are clustered to one
photon--jet system if they are collinear.  In this context the energy
fraction of the photon inside the photon--jet system is defined as
\begin{equation}
\label{eq:photonenergyfraction}
z_{\gamma} = \frac{E_{\gamma}}{E_{\gamma} + E_{\mathrm{jet}}} \, ,
\end{equation}
where $E_{\gamma}$ and $E_{\mr{jet}}$ are the energies of the photon and 
the jet in the collinear regime, respectively. 
We define the photon--jet system as a photon if $z_{\gamma}$ is larger
than a certain cut value $z_{\mr{cut}}$,  otherwise we consider it as
a jet and discard the event.
The complementary approach was used in \citere{Denner:2009gj} to define
$\PW+\mr{jet}$ production including NLO QCD+EW corrections;
there events were only kept if $z_{\gamma}$ was smaller than some
value $z_{\mr{cut}}$.

Based on this procedure a quark-to-photon fragmentation function 
has been measured by ALEPH~\cite{Buskulic:1995au}. 
This fragmentation function 
is used to compensate the collinear singularity (appearing 
if the quark radiates a collinear photon) by
a redefinition of the fragmentation function similarly to the 
redefinition of the PDFs. 
At NNLO $\O (\alphas^{2})$ 
one would even have 
to introduce a gluon-to-photon fragmentation function. 

For extracting the collinear singularity from the real corrections 
the QED dipole subtraction formalism from \citere{Dittmaier:2008md} is applied 
again. Following this method the subtraction function has the 
same structure as the subtraction function for the 
quark--antiquark-induced real EW corrections (\ref{subtractionfunction}):
\begin{equation}
|\M_{\sub} (\Phi_1)|^2 = - \sum_{f,f' \atop f \neq f'} Q_{f} \sigma_{f} Q_{f'} 
\sigma_{f'} e^2 g^{\klam{\sub}}_{ff'} (p_{f},p_{f'},k) \Big| 
\M_{\mathrm{0}} \left( \tilde{\Phi}_{\mathrm{0},ff'}\right) \Big|^2 \, ,
\label{eq:EWsubtraction}
\end{equation}
where $\M_{0}$ is the Born matrix element of the process without photon radiation. 
Actually the sum over the emitter $f$ runs over all charged fermions
in the IS or FS of the process. 
However, demanding a resolved hard photon with non-vanishing $p_{\mathrm{T}}$
and separated from the FS lepton,
effectively reduces this
sum to the term where $f$ is a FS quark or antiquark $i$ appearing in the real
NLO corrections. Photons collinear to IS partons are excluded by the photon
acceptance cuts, photons collinear to FS leptons are excluded by dedicated 
separation cuts. 
The subtraction function, thus, reads
\begin{equation}
|\M_{\sub} (\Phi_1)|^2 = - \sum_{f' \neq i} Q_{i} \sigma_{i} Q_{f'} 
\sigma_{f'} e^2 g^{\klam{\sub}}_{if'} (p_{i},p_{f'},k) \Big| 
\M^{\PW+\mr{jet}}_{0} \left( \tilde{\Phi}_{\mathrm{0},if'}\right) \Big|^2 \, ,
\end{equation}
where $i$ denotes an (anti-)up- or down-type quark and the matrix
element $\M^{\PW+\mr{jet}}_{0}$ in (\ref{eq:fragsubtraction}) is the
Born amplitude for $\PW + \mr{jet}$ production without a photon in the
final state.  
Note also that identifying a collinear photon--jet pair as a photon only if
$z_\gamma>z_{\cut}$ excludes the soft-photon singularity from the phase
space relevant for $\PW+\gamma$ production, i.e.\ only the collinear
asymptotics of the dipole functions is required to obtain finite integrals.
Since no soft singularity needs to be subtracted and
all dipoles have the same collinear limit, they can be
set equal (up to their charge factors), i.e.\
one specific emitter--spectator pair can be chosen.
Furthermore, charge conservation can be used to eliminate the
remaining sum $\sum_{f'\neq i}\sigma_{f'}Q_{f'}=-\sigma_{i}Q_{i}$.
Choosing the charged FS lepton $l$ as the spectator the subtraction
function reads
\begin{equation}
|\M_{\sub} (\Phi_1)|^2 = Q_{i}^2e^2g^{(\sub)}_{il} (p_{i},p_{l},k) \,\Bigl| 
\M^{\PW+\mr{jet}}_{0} \left(\tilde{p}_{i},\tilde{p}_{l},\{\tilde{k}_{n}\}\right) \Bigr|^2 \, .
\label{eq:fragsubtraction}
\end{equation}
The quantities
$\tilde{p}_{i},\tilde{p}_{l}$, and $\{\tilde{k}_{n}\}$ represent the
momenta of the projected phase space $\tilde{\Phi}_{\mathrm{0}}$.
Since in our case only a FS spectator appears, no boost has to be
applied and the momenta $\{\tilde{k}_{n}\}$ equal the momenta
$\{k_{n}\}$ of the $(N+1)$-particle phase space. The $\PW + \mr{jet}$
Born processes are
\begin{align}
\Pu_i \,\Pg      &\to \Plp \Pnl \, \Pd_j \, , \nonumber \\
\Pdbar_j \, \Pg   &\to \Plp \Pnl \, \Pubar_i 
\label{eq:wjetborngluon}
\end{align}
with an IS gluon and 
\begin{align}
\Pu_i \,\gamma      &\to \Plp \Pnl \, \Pd_j \, , \nonumber \\
\Pdbar_j \, \gamma   &\to \Plp \Pnl \, \Pubar_i 
\label{eq:wjetbornphoton}
\end{align}
with an IS photon.  The dipole subtraction function for the case of
massless FS emitter and FS spectator from (\ref{eq:fragsubtraction})
is given by
\begin{align}
g^{(\sub)}_{il} &= \frac{1}{(p_{i} k)(1 - y_{il})} 
\left[ \frac{2}{1 - z_{il} (1 - y_{il})} - 1 - z_{il} \right] \, , 
\label{eq:fragdipoles}
\end{align}
where $y_{il}$ and $z_{il}$ are defined as follows,
\begin{align}
  y_{il} = \frac{p_{i}k}{p_{i}p_{l}+p_{i}k+p_{l}k}  \,\, , \qquad
  z_{il} = \frac{p_{i}p_{l}}{p_{i}p_{l}+p_{l}k} = 1-z_{\gamma l}\,.
\end{align}
In the collinear limit $(p_{i}k) \rightarrow 0$ they behave as
$y_{il} \rightarrow 0$, $z_{il} \rightarrow z_{i}$, and $z_{\gamma l}
\rightarrow z_{\gamma}$, 
where $z_{i}$ and $z_\ga$ are the energy fractions in the
quark--photon system in the collinear limit,
\begin{align}
  z_{i} = \frac{E_{i}}{E_{i}+E_{\gamma}}=1-z_{\gamma}\,.
\end{align} 
Otherwise, the subtraction procedure works analogously as
for the real EW corrections in the NCS case,
\begin{equation}
\int \, \rd \Phi_1 \biggl[|\M_{\mr{real}}|^2 \Theta_{\cut} 
(p_{i},k,p_{l},\{k_n\}) 
- |\M_{\sub}|^2 \Theta_{\cut} \left( (1-z_{\gamma l}) 
\tilde{p}_{i}, z_{\gamma l} \tilde{p}_{i}, 
\tilde{p}_{l}, \{\tilde{k}_n\} \right) \biggr] \, ,
\label{eq:NCSfrag}
\end{equation}
with the only difference that we make the dependence on $z_{\gamma l}$ 
explicit instead of $z_{il}$, since we want to cut on $z_{\gamma l}$. 
Following the calculations in \citeres{Dittmaier:2008md,Denner:2010ia}, we find 
for the integrated dipole contribution
to the partonic cross section
\begin{align}
\md\hat\sigma_{\mr{sub}}(z_{\mr{cut}},m_{i})  =&\,\, 
\frac{\alpha Q_{i}^2}{4 \pi\hat s} \,\rd \tilde{\Phi}_{\mathrm{0}} 
\int_{z_{\cut}}^{1} \! \rd z_{\gamma} \, 
{\cal \bar{G}}^{(\sub)}_{i l} (P^2_{i l},1-z_{\gamma}) \nonumber\\
&\times \left| \M^{W+\jet}_{\mathrm{0}} (\tilde{p}_{i} , \tilde{p}_{l},\{\tilde{k}_{n}\}) 
\right|^2 \Theta_{\cut} \left( p_i=(1-z_{\gamma}) 
\tilde{p}_{i}, k=z_{\gamma} \tilde{p}_{i}, \tilde{p}_{l}, \{\tilde{k}_n\} \right) \, ,
\label{eq:fragreadded}
\end{align}
where $z_{\mr{cut}}$ is the lower limit on $z_{\gamma}$ and
$\sqrt{\hat s}$ is the centre-of-mass energy of the partonic scattering process.
Note that the soft-singular endpoint appearing at $z_\ga\to0$ is excluded
by the cut $z_\ga>z_{\mr{cut}}$, so that we need not introduce a plus distribution to
isolate a soft singularity.
The function ${\cal \bar{G}}^{(\sub)}$ for FS emitter 
and FS spectator reads
\begin{align}
{\cal \bar{G}}^{(\sub)}_{i l} (P^2_{i l},z_{i}) &= P_{ff} (z_{i}) 
\left[ \ln \left( \frac{P^2_{i l} z_{i}}{m^2_{i}} \right)  - 1 
\right] + (1-z_{i}) \ln (1-z_{i}) + \klam{1 -z_{i}} \, ,
\label{eq:finfinncsfunctions}
\end{align}
where $P^2_{i l}$ is the squared invariant mass of the FS 
quark and the charged lepton, and $m_{i}$ is the quark mass 
regularizing the collinear singularity. 

The subtraction procedure outlined above does not remove the FS singularity
appearing in collinear photon--jet pairs, it only isolates the corresponding
divergence in terms of a logarithm $\ln m_i$ of the light-quark mass $m_i$ 
(or alternatively a dimensionally regularized divergence).
In order to restore IR safety the situation where a 
FS quark fragments into a collinear quark--photon pair 
has to be considered. The occurring mass singularity
can be treated by introducing a 
quark-to-photon fragmentation function $D_{q \rightarrow 
\gamma}(z_{\gamma})$ which is defined in 
\citeres{Glover:1993qtp,Glover:1994th} and was measured in 
\citere{Buskulic:1995au}. It describes 
the probability that a quark fragments into a photon with the 
energy fraction $z_{\gamma}$. As worked out in \citere{Glover:1994th}, 
the fragmentation contribution $\hat\sigma_{\mr{frag}}(z_{\mr{cut}})$ to 
the partonic cross section is given by
\begin{equation}
\md\hat\sigma_{\mr{frag}} (z_{\mr{cut}}) = \md\hat\sigma_{0}\int^{1}_{z_{\mr{cut}}} \rd z_{\gamma} \,
D^{\mr{bare}}_{q \rightarrow\gamma}(z_{\gamma}) \, ,
\label{eq:csfragfunction}
\end{equation}
where $\md\hat\sigma_{0}$ denotes the partonic LO cross section.
The collinear singularity can now be compensated by a redefinition 
of the bare fragmentation function $D^{\mr{bare}}_{q \rightarrow\gamma}$. 
We split off the singular contribution regularized  by 
the infinitesimal quark mass $m_i$, which introduces a dependence on the 
factorization scale $\mu_{\mr{F}}$ separating the perturbative from the
non-perturbative region \cite{Denner:2010ia},
\begin{equation}
D_{q \rightarrow \gamma}^{\mathrm{bare,MR}} (z_{\gamma}) = \frac{\alpha 
Q^{2}_{i}}{2 \pi} P_{ff} (1-z_{\gamma}) \left( \ln 
\frac{m_i^2}{\mu_{\mathrm{F}}^2} + 2 \ln z_{\gamma} + 1 \right) + 
D_{q \rightarrow \gamma}^{\overline{\mathrm{MS}}} (z_{\gamma},\mu_{\mathrm{F}}) \, ,
\label{eq:fragfunction}
\end{equation}
where the label ``MR'' stands for the employed mass regularization.
As shown in \citere{Denner:2010ia}, the finite contribution 
$D_{q \rightarrow \gamma}^{\overline{\mathrm{MS}}}$ defined in this way is
equivalent to the standard $\overline{\mathrm{MS}}$~scheme of dimensional
regularization.
Here, we employ
the parametrization used by the ALEPH collaboration
\cite{Buskulic:1995au}
\begin{equation}
D_{q \rightarrow \gamma}^{\mathrm{ALEPH},\overline{\mathrm{MS}}} 
(z_{\gamma},\mu_{\mathrm{F}}) = \frac{\alpha Q_i^2}{2 \pi} \left( 
P_{ff} (1-z_{\gamma}) \ln 
\frac{\mu_{\mathrm{F}}^2}{(1-z_{\gamma})^2 \mu_0^2} + C \right) 
\, . 
\label{eq:ALEPHfragfunction}
\end{equation}
The fit parameters $\mu_0^2$ and $C$ constrained by 
$C = -1 + \ln (2 \mu_0^2/\MZ^2)$ are
\begin{equation}
\mu_0 = 0.14 \GeV \text{  and  } C = -13.26 \, . 
\label{eq:fitparametersfargfunction}
\end{equation}
Combining (\ref{eq:fragreadded}) and 
(\ref{eq:fragfunction}), setting $1-z_{\gamma}=z_{i}$
and performing the $z_{i}$-integration 
analytically leads to (see Eq.~(4.63) in \citere{Denner:2010ia})
\begin{align}
\md\hat\sigma_{\mr{sub}}(z_{\mr{cut}},m_{i}) &
+ \md\hat\sigma_{\mr{frag}}(z_{\mr{cut}},m_{i}) \nonumber\\ 
& =
 \frac{\alpha Q^{2}_{i}}{4 \pi\hat s}\, \rd \tilde{\Phi}_{\mathrm{0}} \left| 
\M^{W+\jet}_{\mathrm{0}} (\tilde{\Phi}_{\mathrm{0}}) \right|^{2} \int_{0}^{1-z_{\cut}} \rd 
z_{i} \left( D_{q \rightarrow \gamma}^{\mathrm{bare,MR}} (1-z_{i}) + 
\bar{\cal{G}}^{(\sub)}_{il} (P^2_{il},z_{i}) \right)  \nonumber \\
& = \frac{\alpha Q_{i}^{2}}{4 \pi\hat s}\, \rd \tilde{\Phi}_{\mathrm{0}} \left| 
\M^{W+\jet}_{\mathrm{0}} (\tilde{\Phi}_{\mathrm{0}}) \right|^2 
\bigg\{ \left( 1 + C + \frac{z_{\cut}}{2} \right) (1 - z_{\cut}) \nonumber \\
& \qquad - \left( \frac{1}{2} (1-z_{\cut})(3-z_{\cut}) + 
2 \ln (z_{\cut}) \right) \ln \left( \frac{z_{\cut}}{1-z_{\cut}} 
\frac{P^2_{il}}{\mu_0^2} \right)   \nonumber \\
& \qquad + 2 \Li (1-z_{\cut}) + \frac{3}{2} \ln (z_{\cut}) \bigg\}\, ,
\label{eq:totalfragcontribution}
\end{align}
which is finite and only depends on the value $z_{\cut}$.

\section{Numerical results}
\label{se:numres}

\subsection{Input parameters and setup}
\label{se:SMinput}

The relevant SM input parameters are
\begin{equation}\arraycolsep 2pt
\begin{array}[b]{lcllcllcl}
\GF & = & 1.1663787 \times 10^{-5} \GeV^{-2}, \quad&
\alpha(0) &=& 1/137.035999074\,,&
\alpha_{\mathrm{s}}(\MZ) &=& 0.119\, , 
\\
M_\PH &=& 125\GeV, &
m_\mu &=& 105.6583715\MeV, &
m_\Pt &=& 173.07\;\GeV,
\\
\MW^{\OS} & = & 80.385\GeV, &
\Gamma_\PW^{\OS} & = & 2.085\GeV, \\
\MZ^{\OS} & = & 91.1876\GeV, &
\Gamma_\PZ^{\OS} & = & 2.4952\GeV, &
 \\
|V_{\rm\Pu\Ps}| & = & |V_{\rm\Pc\Pd}| = 0.225\,, &
|V_{\rm\Pu\Pd}| & = & |V_{\rm\Pc\Ps}| = \sqrt{1 - |V_{\rm\Pu\Ps}|^2}\,. &
\end{array}
\label{eq:SMpar}
\end{equation}
All parameters but $\alpha_{\mathrm{s}}(\MZ)$, which is provided by
the PDF set, are extracted from \citere{Beringer:1900zz}. 
The masses of all quarks but the top quark are set to zero.
CKM mixing between the
first two quark generations is
taken into account in all partonic cross sections, but
mixing to the third generation is not included, since it is negligible.
This implies that there is no contribution from bottom quarks in the
initial state and that the CKM matrix drops out in the flavour sum
of closed fermion loops.
Thus, the CKM matrix factorizes from all amplitudes, so that
only one generic amplitude has to be evaluated when convoluting the
squared matrix elements with the PDFs.

Owing to the presence of an on-shell external photon, we always take
one electromagnetic coupling constant $\al$ at zero momentum transfer,
$\al=\al(0)$.  For the other couplings, e.g.\ of the $\PW$~boson to
fermions, we determine the electromagnetic coupling constant in the
$G_{\mu}$ scheme, where $\alpha$ is defined in terms of the Fermi
constant,
\begin{align}
   \al_{G_{\mu}}=\frac{\sqrt{2}}{\pi}G_{\mu}M^{2}_{\PW}\klam{1-\frac{M^{2}_{\PW}}{M^{2}_{\PZ}}}.
\end{align}
This definition effectively resums contributions associated with the
evolution of $\al$ from zero momentum transfer to the electroweak
scale and includes universal corrections to the $\rho$-parameter.  In
this scheme large fermion-mass logarithms are effectively resummed
leading to an independence of logarithms of the light fermion
masses \cite{Denner:2005fg}
(see also the discussion in the ``EW dictionary'' in \citere{Butterworth:2014efa}).
Using this mixed scheme the squared LO
amplitude is proportional to $\alpha(0)\alpha^{2}_{G_{\mu}}$.
In the relative EW corrections we set the additional coupling factor $\alpha$
to $\al_{G_{\mu}}$, because this coupling is adequate for the most pronounced
EW corrections which are caused by soft/collinear weak gauge-boson exchange
at high energies (EW Sudakov logarithms, etc.).

We apply the complex-mass scheme \cite{Denner:1999gp,Denner:2005fg,Denner:2006ic} to describe the 
W-boson resonance by introducing complex vector-boson masses according to
\begin{align}
M^{2}_{\PW}  \rightarrow \mu^{2}_{\PW} = M^{2}_{\PW}-\ri M_{\PW}\Gamma_{\PW}\,, \qquad
M^{2}_{\PZ}  \rightarrow \mu^{2}_{\PZ} = M^{2}_{\PZ}-\ri M_{\PZ}\Gamma_{\PZ}
\end{align}
with constant widths.  However, at LEP and the Tevatron the on-shell
(OS) masses of the vector bosons were measured, which correspond to
running widths. The OS masses $M^{\OS}_{\PW}$, $M^{\OS}_{\PZ}$ and
widths $\Gamma^{\OS}_{\PW}$, $\Gamma^{\OS}_{\PZ}$ have to be
converted to the pole values using the relations
\cite{Bardin:1988xt}
\begin{align}
  M_{\mathrm{V}}=M^{\OS}_{\mathrm{V}}/\sqrt{1+\klam{\Gamma^{\OS}_{\mathrm{V}}/M^{\OS}_{\mathrm{V}}}^{2}}, 
     \qquad \Gamma_{\mathrm{V}}=
            \Gamma^{\OS}_{\mathrm{V}}/\sqrt{1+\klam{\Gamma^{\OS}_{\mathrm{V}}/M^{\OS}_{\mathrm{V}}}^{2}}
     \qquad (\mathrm{V}=\mathrm{W},\;\PZ)\,,
\end{align}
leading to 
\begin{align}
  \begin{array}[b]{r@{\,}l@{\qquad}r@{\,}l}
\MW &= 80.3580\ldots\GeV, & \Gamma_{\PW} &= 2.0843\ldots\GeV, \\
\MZ &= 91.1535\ldots\GeV,& \Gamma_{\PZ} &= 2.4943\ldots\GeV.
\label{eq:m_ga_pole_num}
\end{array}
\end{align}

Calculating the hadronic cross section, we use the 
${\cal O}(\alpha)$-corrected NLO PDF set NNPDF23~\cite{Ball:2013hta}, which
includes the two-loop running of $\al_{s}$ for five active flavours
($n_{f}=5$).

The factorization and the renormalization scales $\mu_{\mathrm{F}}, \mu_{\mathrm{R}}$
are set equal
throughout our calculation.
Following \citeres{Dixon:1999di,Haywood:1999qg}, 
we choose the scales as
\begin{align}
  \mu^{2}_{\mathrm{F}}=\mu^{2}_{\mathrm{R}}=\frac{1}{2}\klam{\MW^{2}+p^{2}_{\mathrm{T},\PW}
          +p_{\mathrm{T},\gamma_{1}}^{2}+p_{\mathrm{T},\gamma_{2} / \mathrm{jet}}^{2}},
\end{align}
where $p_{\mathrm{T},\PW}$ is the transverse momentum of the massive vector boson defined by
\begin{align}
  p_{\mathrm{T},\PW} = |{\bf p}_{\mathrm{T},l}+{\bf p}_{\mathrm{T},\nu}|,
\end{align}
and $p_{\mathrm{T},a}=|{\bf p}_{\mathrm{T},a}|$ denotes the absolute value
of the transverse three-momentum ${\bf p}_{\mathrm{T},a}$ of particle $a$.
The photons $\ga_1$ and $\ga_2$ are assigned so that $p_{\mathrm{T},\gamma_{1}}>p_{\mathrm{T},\gamma_{2}}$.
In LO the transverse momenta $p_{\mathrm{T},\gamma_{2} / \mathrm{jet}}$ are zero.

\subsection{Phase-space cuts and event selection}
\label{se:cuts}
The process $\Pp\Pp\to\PW^+ + \gamma \to \Pl^+\nu_\Pl + \gamma + X$
requires the recombination of FS photons with FS partons and, where
appropriate, of FS photons with charged leptons in regimes of phase space where photon and parton/lepton
are collinear.  Furthermore, we impose several cuts to account for the
detector acceptance.  The phase-space cuts and the event selection are
inspired by the recent ATLAS and CMS papers
\cite{Aad:2013izg,Chatrchyan:2013fya,Aad:2014fha} 
analysing $V\gamma$ final states.

\subsubsection{Recombination}
To decide whether a photon and a FS particle need to be recombined we
use the Euclidean distance in the $y$--$\phi$ plane,
$R_{ij}=\sqrt{\klam{y_{i}-y_{j}}^{2}+\phi^{2}_{ij}}$, where 
$y=\frac{1}{2}\ln\left[\klam{E+p_{\mathrm{L}}}/\klam{E-p_{\mathrm{L}}}\right]$
denotes the rapidity.
In this equation $E$ is the energy and $p_{\mathrm{L}}$ the
longitudinal momentum of the respective particle with respect to the
beam axis. The value $\phi_{ij}$ denotes the angle between the
particles $i$ and $j$ in the plane perpendicular to the beams.  The
recombination proceeds as follows:

\begin{enumerate}
\item A photon and a charged lepton are never recombined if we
  consider ``bare'' muons.  Otherwise recombination is applied if
  $R_{l^+\gamma}<0.1$, which means that their four-momenta are added.
  In case of two photons in the final state, the recombination is first done
  with the photon that yields the smaller $R_{l^+\gamma}$, then this condition
  is checked for the second photon.
\item Two photons are recombined if $R_{\gamma\gamma}<0.1$.
\item A photon and a jet are recombined if the distance between them
  becomes $R_{\gamma\mathrm{jet}}<0.5$. After recombining them, the
  energy fraction
  $z_{\gamma}=E_{\gamma}/\klam{E_{\gamma}+E_{\mathrm{jet}}}$ of the
  photon inside the photon--jet system
  is determined.
  If $z_{\gamma}$ is smaller than the cut value $z_{\mr{cut}}=0.9$ the
  event is regarded as a part of the process $\PW+\mathrm{jet}$ and is
  therefore rejected.
\end{enumerate}
The case where more than two particles are recombined is excluded by
our basic cuts.  Results are presented for ``bare'' muons and for
photon recombination with leptons. The latter results 
hold for electrons as well as for muons, since the
lepton-mass logarithms cancel as dictated by the KLN theorem
\cite{Kinoshita:1962ur,Lee:1964is}.

\subsubsection{Basic cuts}
\label{sec:basic-cuts}
After recombination, we define $\PW+\gamma$ events by the following
cut procedure:
\begin{enumerate}
\item We demand a charged lepton with transverse momentum 
  $p_{\mathrm{T},l}>25 \;\mathrm{GeV}$ and missing
  transverse momentum $\slashed{p}_{\mathrm{T}}>25 \;\mathrm{GeV}$,
  where $\slashed{p}_{\mr{T}}$ is equal to the neutrino transverse momentum.
\item We demand at least one photon with transverse momentum
  $p_{\mathrm{T},\gamma}>15 \;\mathrm{GeV}$ that is isolated from the
  charged lepton with a distance of $R_{l^+\gamma}>0.7$.
\item The charged lepton and 
  the photon passing the cuts at step 2
  have to be central, i.e.\ their rapidities have to be in the range
  $\abs{y}<2.5$.
\item 
Only events with a transverse mass of the lepton pair 
$\MtW>40\;\GeV$ are accepted.
\end{enumerate} 
We present results with and without applying a jet veto. 
Applying a jet veto means
that all events including a FS jet with $p_{\mr{T,\mathrm{jet}}}>100\GeV$
are discarded.

\subsection{Results on total cross sections}
\label{se:CSresults}
   
\begin{table}
  \centering\renewcommand{\arraystretch}{1.2}
  \begin{tabular}[H]{c|rrr}
    \multicolumn{4}{c}{$\Pp\Pp \to \Plp \Pnl\; \gamma + \mathrm{X} $} \\
    \hline\hline
    $\sqrt{s}/\TeV\; $ & $7$\phz & $8$\phz & $14$\phz \\
    \hline\hline
     $\si^{\mr{LO}}             /\fba\;$ &$\;846.53(4)$&$\;940.30(4)$&$\;1447.99(4)$  \\
    \hline \hline
    $\de_{\EW, \Pq\overline{\Pq}}^{\mr{NCS}}/\% \;$   
    &$\; -3.15$\phz &$\; -3.14$\phz &  $\; -3.15$\phz     \\
\hline 
    $\de_{\EW, \Pq\overline{\Pq}}^{\mathrm{CS}}/\% \;$    
    &$\; -1.95$\phz &$\; -1.94$\phz & $\; -1.95$\phz      \\ 
    \hline \hline
    $\de_{\EW, \Pq\ga}/\% \;$        &$\; 1.04$\phz &$\; 1.10$\phz &$\; 1.30$\phz         \\ 
    \hline 
    $\de^{\mr{veto}}_{\EW, \Pq\ga}/\% \;$ &$\; 0.74$\phz &$\; 0.76$\phz &$\; 0.84$\phz         \\ 
    \hline \hline
    $\de_{\QCD}/\% \;$                &$\; 122.33(4)$ &$\; 128.30(5)$ &$\; 153.61(3)$    \\ 
    \hline
    $\de^{\veto}_{\QCD}/\% \;$           &$\; 112.47(4)$ &$\; 117.15(5)$ &$\; 135.97(9)$     \\ 
    \hline \hline
  \end{tabular}
  \caption{\label{ta:totcs} 
    Integrated cross sections  and relative corrections for different
    LHC energies.  The EW corrections to the quark--antiquark 
    annihilation channels are provided with (CS) and  without (NCS)
    lepton--photon recombination. EW corrections from the
    photon-induced channels and QCD corrections are shown with a jet
    veto (veto) as well as without a jet veto. The numbers in
    parentheses denote the integration error in the last digit. This
    error is negligible for the relative corrections at the given accuracy.} 
\end{table}

In \refta{ta:totcs} we present the LO cross sections $\si^{\mr{LO}}$
for different $\Pp\Pp$ 
centre-of-mass energies $\sqrt{s}$ and the different types of
relative corrections $\de$ defined in \refeq{eq:relcor}.  For the EW
corrections resulting from the quark--antiquark channels we show
results for CS and NCS observables. Results for the EW corrections
originating from photon-induced channels and for the QCD corrections
are listed with and without a jet veto.  The different relative
corrections depend only weakly on the collider energy.  By far the
largest effect ($\sim120{-}150\%$) comes from the QCD corrections,
even in case of a jet veto.  About two thirds of the relative QCD
corrections are due to
gluon-induced channels.  The EW corrections to the quark--antiquark
channels are about $-2\%$ and $-3\%$ for the CS and the NCS case,
respectively.  The photon-induced corrections contribute between
$0.7\%$ and $1.3\%$ with and without a jet veto.  Summing up, the
total EW corrections to the integrated cross section are small and not
significant for the most-recent experimental cross-section
measurements. However, larger corrections show up in differential
distributions, as demonstrated in the next section.

\subsection{Results on transverse-momentum and transverse-mass distributions}
\label{se:results_mom_distr}

In the following we present 
the distributions including EW and QCD corrections
for various observables in separate plots.
For each distribution we also show the relative EW corrections of the
$q\bar{q}$ and $q\gamma$ channels as well as the QCD corrections with
and without a jet veto. 
\begin{figure}     
\centerline{
        \includegraphics[scale=0.6]{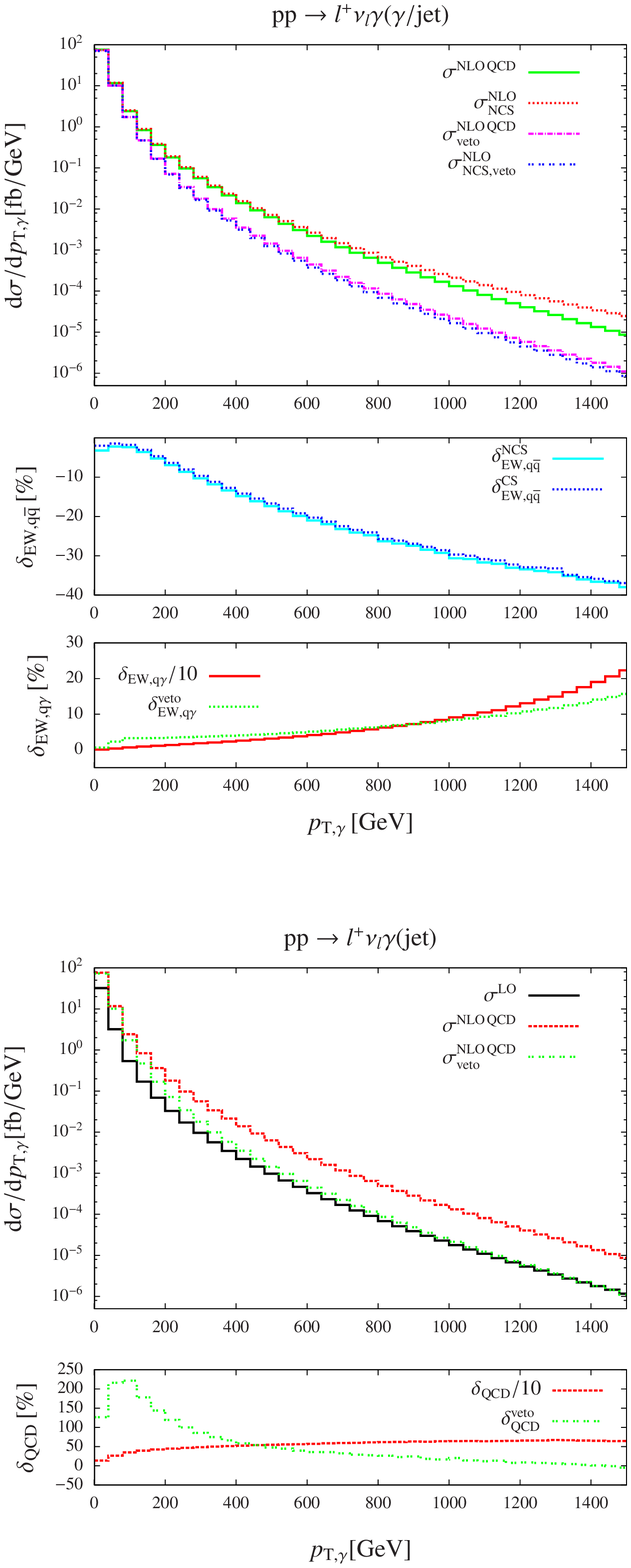}
        \label{Bild3}
\hspace{-3em}
        \includegraphics[scale=0.6]{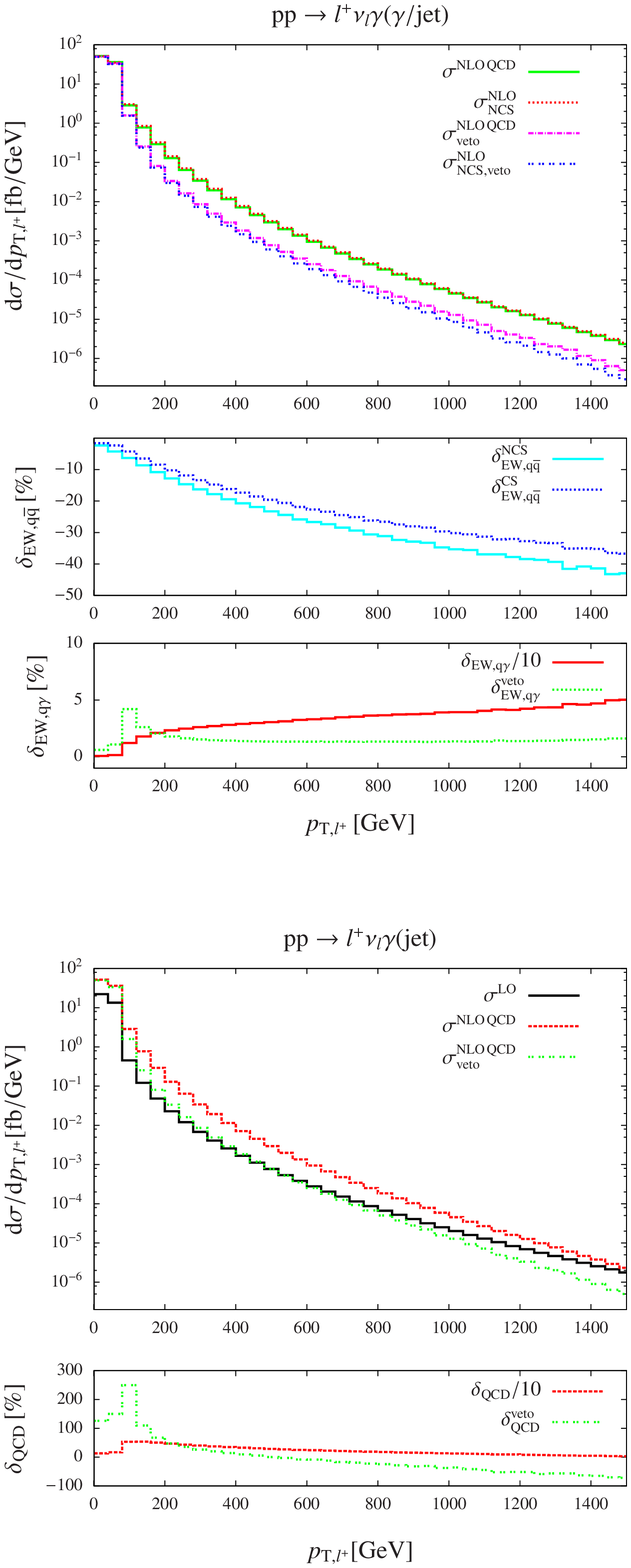}
        \label{Bild4}
}
\caption{\label{fi:pt} Distributions in the
  transverse momentum $p_{\mathrm{T}}$ of the photon (left) and the charged lepton
  (right), including EW (top) and QCD corrections (bottom). For all corrections 
   absolute (upper box) and relative corrections (lower boxes) are shown.}
\end{figure} 


In \reffi{fi:pt} we show results for transverse-momentum
distributions.  Focusing on the $p_{\mathrm{T},\gamma}$~distribution
of the photon (within cuts) 
with the highest transverse momentum we can see (\reffi{fi:pt},
left, bottom) that the QCD corrections 
without a jet veto reach $650\%$ for large
transverse momenta (scaled down by a factor 10 in \reffi{fi:pt}).
This is due to the fact that new production channels occur
(${q}\,\Pg\,\to\,l\,\nu\,\gamma\,{q}$) at NLO QCD causing
large corrections ($550\%$ for large $p_{\mathrm{T},\gamma}$).
However, these large corrections come from events with hard jets,
meaning that they should better be considered as part of $\PW +
\mr{jet}$ rather than $\PW + \gamma$ production. For this reason we
also present results for the case of a jet veto, where all events with
jets with $p_{\mathrm{T},\mr{jet}}>100\GeV$ are discarded.  In this
case the QCD corrections become small for large
$p_{\mathrm{T},\gamma}$, since the jet veto suppresses the
contribution of the real QCD corrections and especially of the
gluon-induced channels.  

Owing to the so-called EW Sudakov logarithms
the EW corrections (\reffi{fi:pt}, left, top) contribute with large 
negative corrections in the
high-$p_{\mathrm{T}}$ range, though one order of magnitude smaller
than the QCD corrections. The CS and the NCS cases hardly differ in
the $p_{\mathrm{T}}$~distribution of the hardest photon, since the
recombination of another photon collinearly emitted off the lepton
only marginally effects the $p_{\mathrm{T}}$ of the hard photon.  
The photon-induced corrections are positive and become surprisingly
large for large transverse momenta, reaching the same order of
magnitude as the QCD corrections.  In fact in the discussion of the
relative impact of the photon-induced corrections it would be more
appropriate to normalize to 
the NLO QCD cross section, which is
dominated by the new channels for hard jet emission at high scales.
With this normalization the $q\gamma$ channels still contribute some
tens of percent at high $p_{\mathrm{T},\gamma}$ with a rising tendency
for growing $p_{\mathrm{T},\gamma}$, which can be understood by the
increasing $\gamma/\Pg$ PDF ratio for high Bjorken-$x$ and the
decrease in the strong coupling constant driven by the dynamical
renormalization scale.  Note, however, 
that the photon PDF at large Bjorken-$x$
suffers from huge uncertainties of up to $100\%$, so that we have to
conclude that the high-$p_{\mathrm{T}}$ tail 
of the $p_{\mathrm{T},\gamma}$
distribution in the TeV range is plagued by PDF uncertainties which
are of the size of the $q\gamma$ contribution itself.
Similarly to the huge QCD corrections,
the large impact of the photon-induced corrections at high $p_{\mathrm{T},\gamma}$
is reduced to the level of $10{-}15\%$ by a jet veto, showing that those large effects
are caused by hard jet emission.
The jet veto, thus, helps to suppress the impact of the $q\gamma$ contribution
and the corresponding large uncertainties in the high-$p_{\mathrm{T}}$ regime. 
After applying the veto, in fact
the quark--antiquark-induced EW corrections become the dominating
corrections for large transverse momenta.

In case of the $p_{\mathrm{T}}$~distribution of the charged lepton
(\reffi{fi:pt}, right) the QCD corrections without jet veto are large
in the small-$p_{\mathrm{T}}$ range and become small for large
transverse momenta. In contrast the EW corrections become sizeable in
the region of large transverse momenta. The corrections are
roughly $5\%$ smaller in the CS case than in the NCS case.  
Collinear photon emission reduces the lepton
momentum, so that events with large $p_{\mathrm{T},\Plp}$ before the emission
migrate to smaller $p_{\mathrm{T},\Plp}$, leading to negative
corrections on the falling distribution in $p_{\mathrm{T},\Plp}$.
Photon recombination damps this effect upon shifting the major part of these
migrating events back to the $p_{\mathrm{T},\Plp}$ value before photon
emission.
For the case without jet veto the
quark--antiquark and photon-induced EW corrections are of
the same order of magnitude, but of opposite sign, and 
accidentally compensate each other to a large extent.
In case of a jet veto the QCD corrections become large and
negative for large $p_{\mathrm{T},\Plp}$. The large negative
corrections result from the quark--antiquark-induced channels, while
the corrections due to the gluon-induced channels remain small
also for a jet veto. This fits well to the fact that the photon-induced
corrections become negligible everywhere.

\begin{figure}
\centerline{
        \includegraphics[scale=0.6]{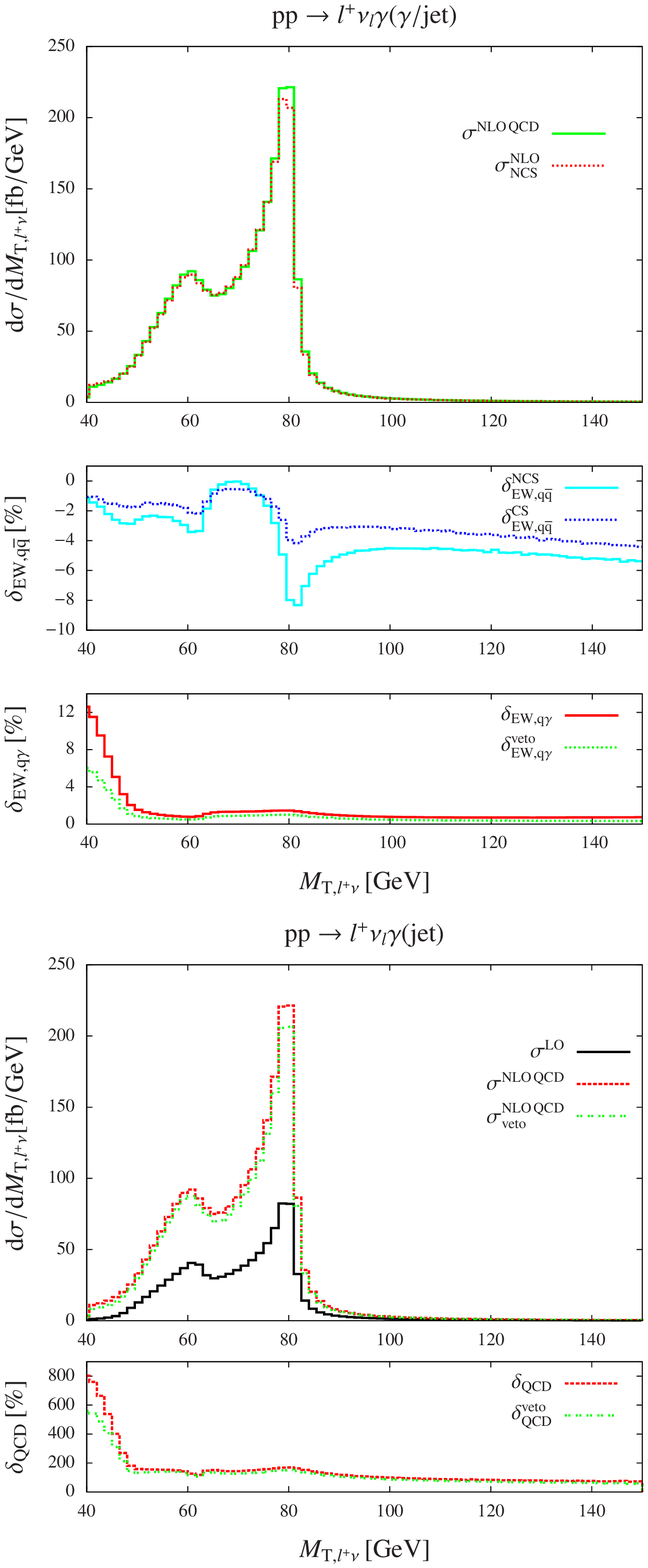}
\hspace{-3em}
        \includegraphics[scale=0.6]{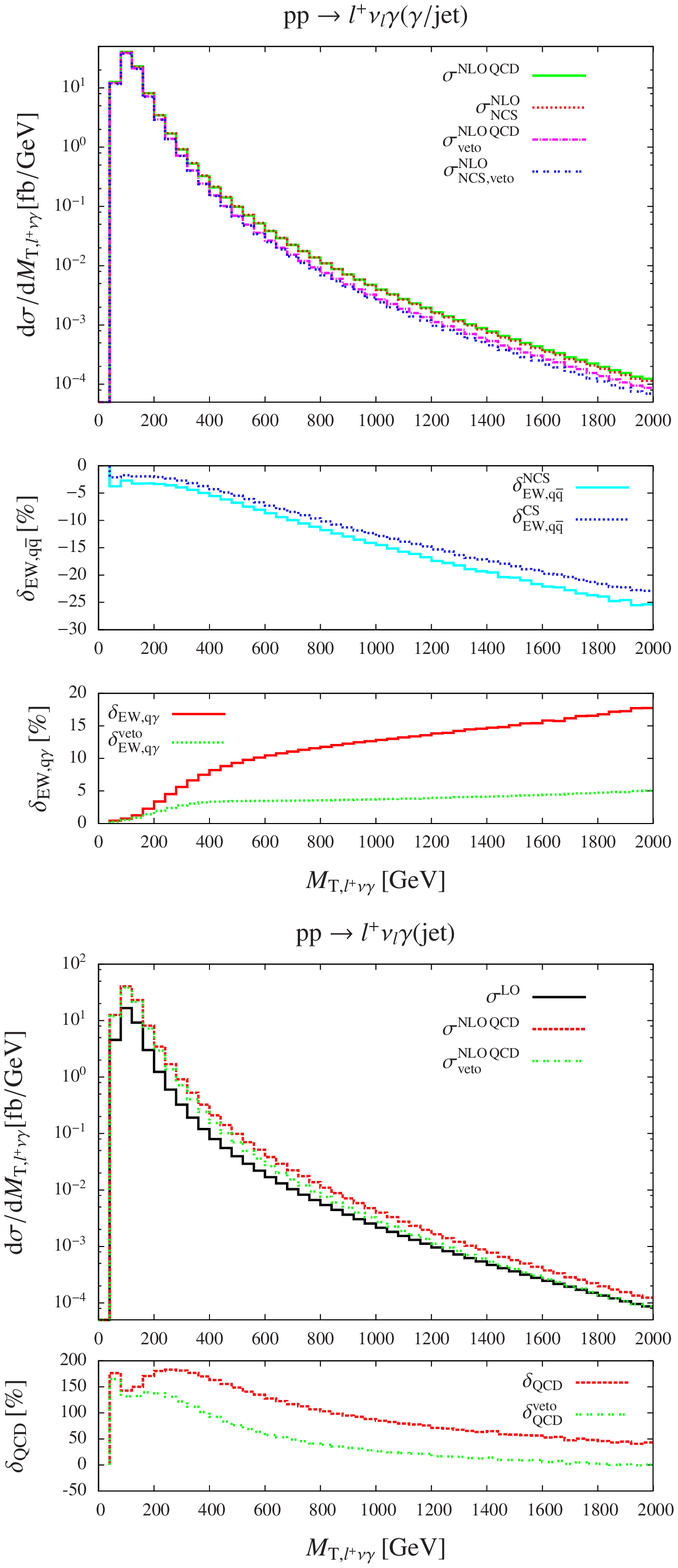}
}
\caption{\label{fi:mt} Distribution in the transverse mass $M_{\mr{T},\Pl^+\nu}$ of the
  charged lepton and neutrino pair (left) and distribution in the transverse
  three-body mass $M_{\mr{T},\Pl^+\nu\gamma}$ of the charged lepton, the
  neutrino, and the hardest photon (right), including EW (top) and QCD
  corrections (bottom). For all corrections 
   absolute (upper box) and relative corrections (lower boxes) are shown.}
\end{figure}

The transverse mass of the W boson and the transverse three-body mass
of the W-decay products and the photon are defined by
\begin{align}
 \MtW & =  \sqrt{2 p_{\mr{T},\Plp} \cdot \slashed{p}_{\mr{T}} 
           \klam{1-\cos{\klam{\Delta\phi_{\Plp,\mr{miss}}}}}}\, ,
\nonumber\\
 \MtWA& =  \sqrt{\klam{\sqrt{M^{2}_{\Plp\gamma}
          +\abs{\textbf{p}_{\mr{T},\Plp} + \textbf{p}_{\mr{T},\gamma}}^{2}}
          +\slashed{p}_{\mr{T}}}^{2}-\abs{\textbf{p}_{\mr{T},\Plp} + \textbf{p}_{\mr{T},\gamma}
          +\slashed{\textbf{p}}_{\mr{T}}}^{2}}\, ,
\end{align}
where $\Delta\phi_{\Plp,\mr{miss}}$ is the azimuthal-angle separation between
the directions of the charged
lepton and the missing transverse momentum,
and $M_{\Pl^+\gamma}$ is the invariant mass of  the
charged lepton and the photon.  The corresponding distributions are shown in
\reffi{fi:mt}. 
The smaller peak in the $\MtW$ distribution at $60\GeV$
appearing already at LO originates from events where the three-body
invariant mass $\MtWA$ lies in the resonance region and the photon is
radiated by the charged FS lepton shifting the peak to
smaller transverse masses.  Since events with photons close to the
FS lepton are discarded, a dip appears above the lower peak.
As can be seen in \reffi{fi:mt}, the QCD corrections are dominating
the $\MtW$ distribution with and without a jet veto.  At the W-mass
peak, the EW corrections reach $-4\%$ with photon recombination and
$-8\%$ in the NCS case, where the photon radiated collinear to the
charged lepton carries away energy, shifting more events to regions of
smaller transverse mass, where those events positively contribute 
to the EW corrections below the W-boson resonance.
The photon-induced corrections are negligible
with and without a jet veto.
We note in passing that previous calculations~\cite{Accomando:2001fn,Accomando:2005ra}
of EW corrections to $\PW+\ga$ production, 
which treat the W~boson in pole approximation, cannot predict the range in
$\MtW$ exceeding $\MW$ which forces the W~boson to go off its mass shell,
while our calculation covers resonant and non-resonant regions in NLO accuracy.

We turn to the $\MtWA$ distribution analysed
experimentally in \citere{Aad:2014fha}. While the QCD corrections are
dominating the region of small transverse masses, the EW and
photon-induced corrections are small and have opposite signs there. In the
high-$\MtWA$ region the situation is different.  Here the QCD
corrections reduce to $50\%$  and in case of a jet veto almost tend to zero.
In contrast, the EW and the photon-induced corrections without a jet
veto are about 20\%, but accidentally compensate each other partly.
However, imposing a jet veto reduces the photon-induced corrections to
$5\%$. As a result, the EW corrections are not compensated by the
photonic corrections anymore, becoming the dominant contribution.

\subsection{Results on rapidity and angular distributions}

\begin{figure}
\centerline{
        \includegraphics[scale=0.6]{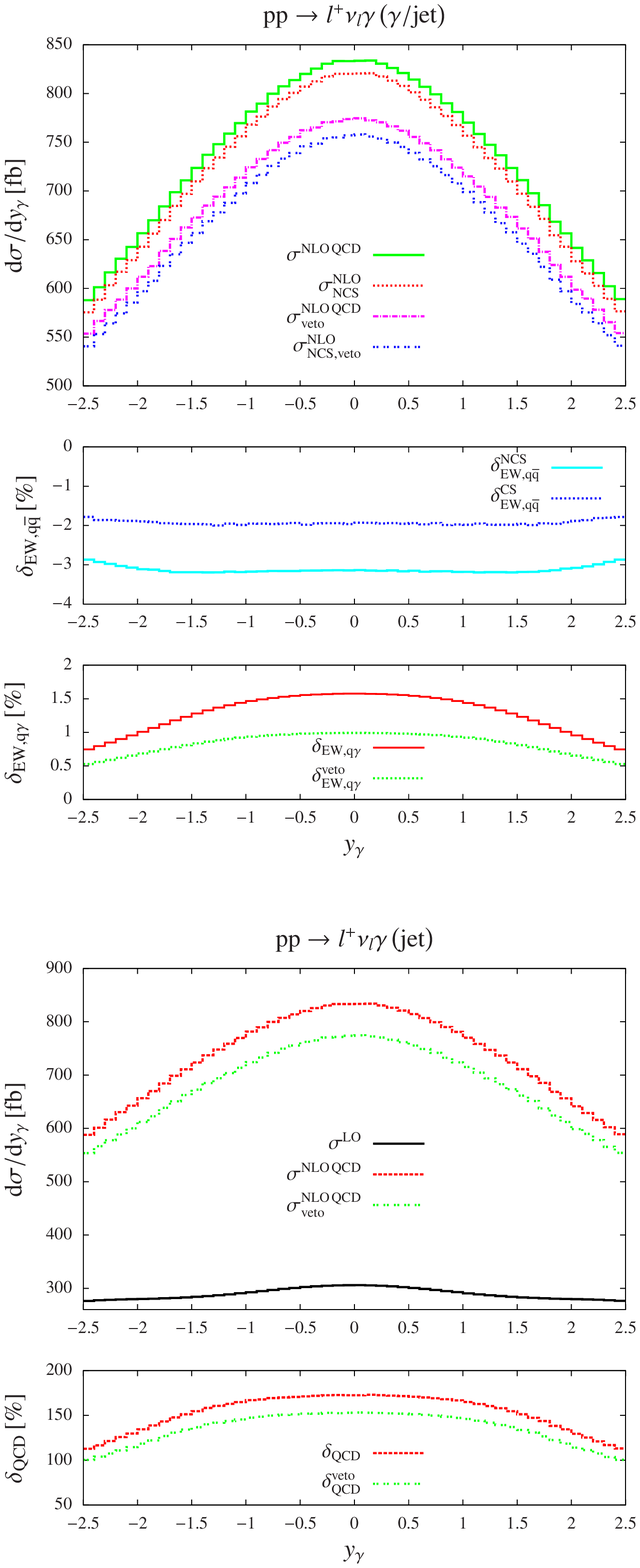}
\hspace{-3em}
        \includegraphics[scale=0.6]{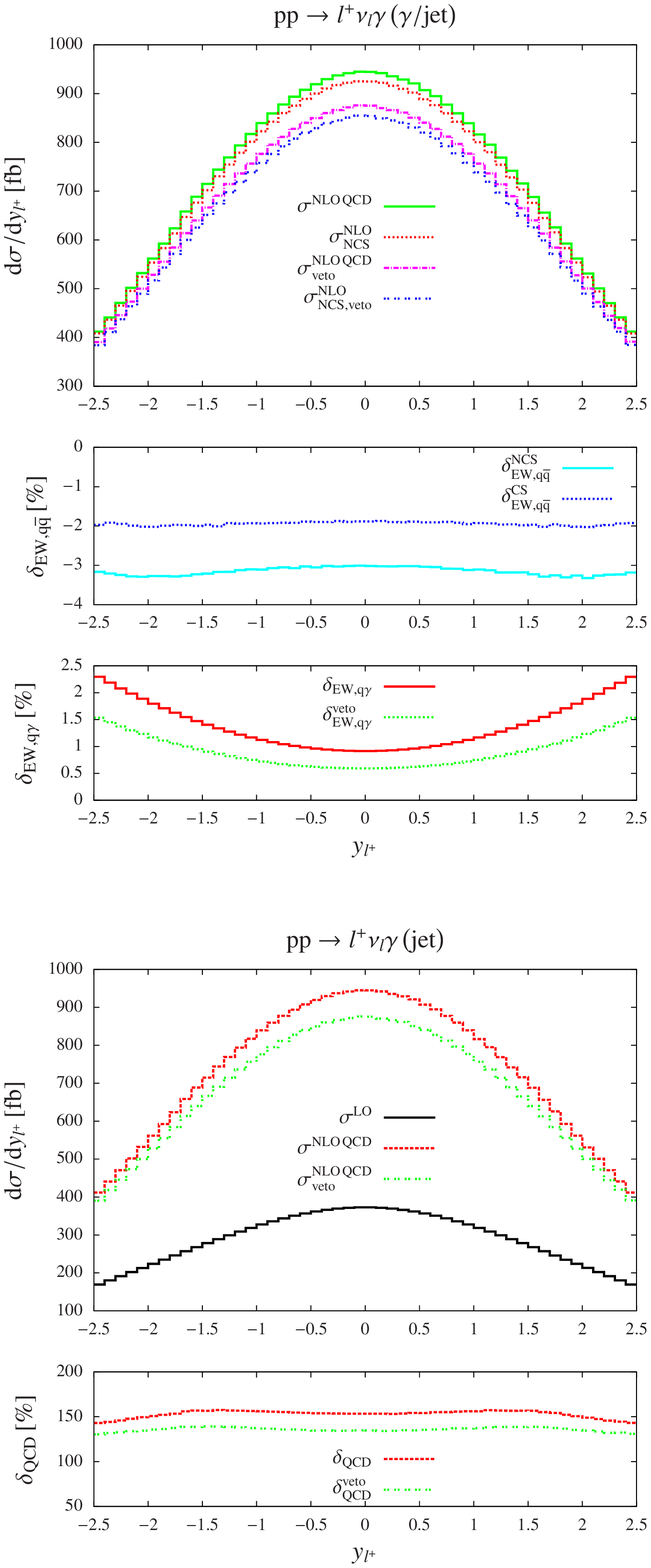}
}
\caption{\label{fi:rap}Distributions in the
  rapidity $y_{\gamma}$ of the photon (left) and the rapidity $y_{\Pl^+}$ of the charged lepton
  (right), including EW (top) and QCD corrections (bottom). For all corrections 
   absolute (upper box) and relative corrections (lower boxes) are shown.}
\end{figure}
\begin{figure}     
\centerline{
        \includegraphics[scale=0.6]{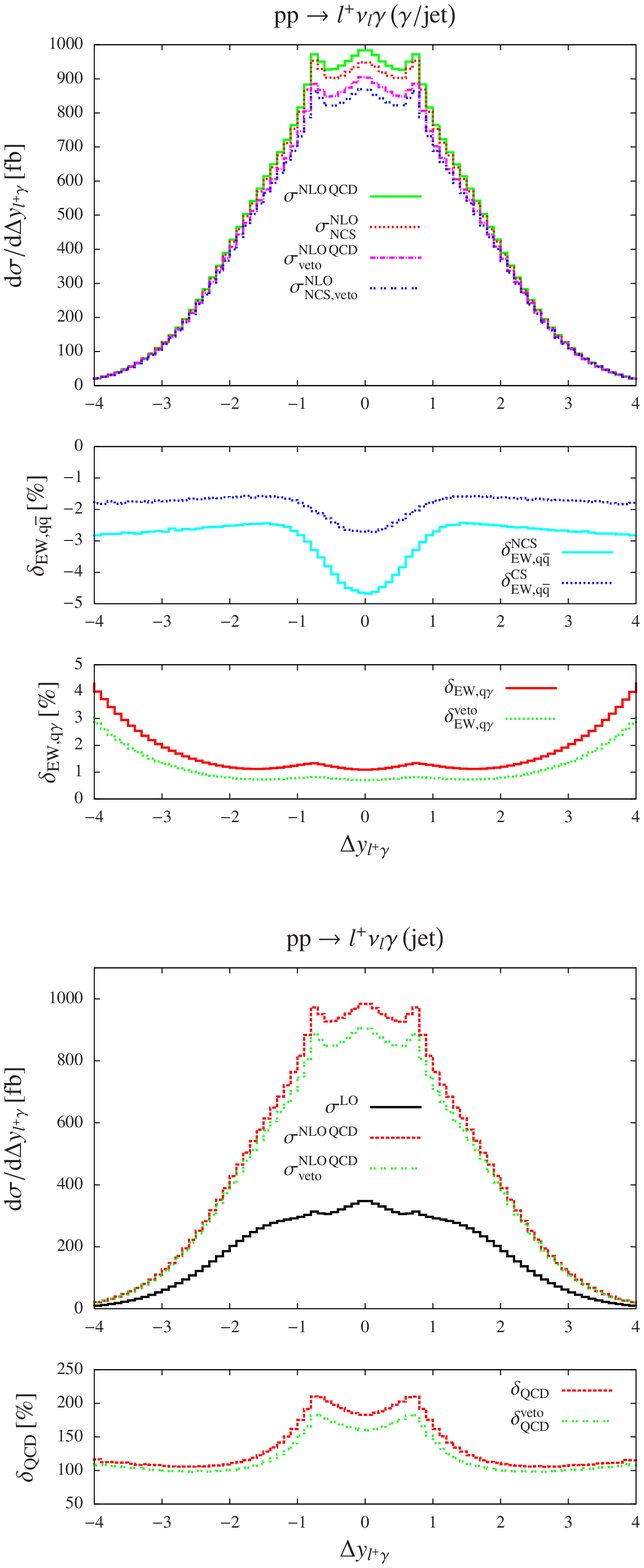}
\hspace{-3em}
        \includegraphics[scale=0.6]{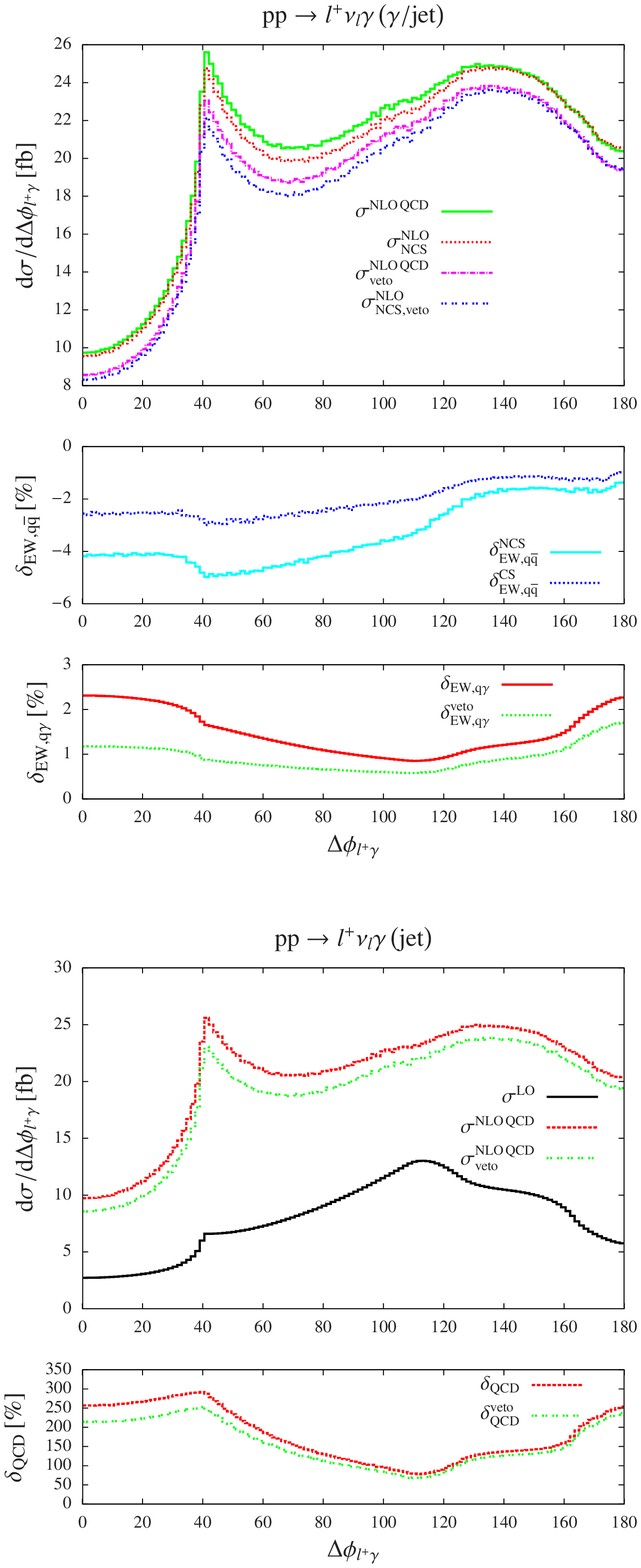}
}
\caption{\label{fi:angle} Distributions in the rapidity difference 
$\Delta y_{\Pl^+\gamma}$ (left) and the azimuthal-angle difference 
$\Delta\phi_{\Pl^+\gamma}$ (right) of the charged lepton and the photon,  
including EW (top) and QCD corrections (bottom). For all corrections 
   absolute (upper box) and relative corrections (lower boxes) are shown.}
\end{figure}
In \reffi{fi:rap} we show the rapidity distributions of the photon and
the charged lepton. In both distributions the relative EW corrections
are small and almost constant over the whole range, and thus
essentially given by the corrections to the total cross section.  The
photon-induced contributions are of comparable magnitude, but have
opposite sign so that they partially cancel the EW corrections. The
QCD corrections amount to $100{-}180\%$ for the rapidity distribution of
the photon and to $140{-}160\%$ for the one of the charged lepton and
lead to sizeable shape distortions.
EW corrections are completely swamped by QCD uncertainties in these
observables.

In \reffi{fi:angle} we present the distributions in rapidity and the
azimuthal-angle difference between the charged lepton and the photon.
Note that the shape of the LO rapidity-difference distribution 
is highly sensitive to the chosen phase-space
cuts.  A potential dip at $\Delta y_{l^+ \gamma}=0$ indicating the
radiation zero \cite{Brown:1979ux,DeFlorian:2000sg,Baur:1994sa} 
is not present in the setup described in
\refse{se:cuts}, but becomes visible for cuts $p_{\mathrm{T}, \gamma} >
20 \GeV$ or $p_{\mathrm{T}, \mathrm{miss}} > 40 \GeV$ (not shown explicitly). 
The kink around
$40 \degree$ in the $\Delta \phi_{l^+ \gamma}$ distribution is a
result of the isolation cut $R_{l^+\gamma}>0.7$ which suppresses the
phase-space region with small azimuthal angle between charged lepton
and photon. The EW and photon-induced corrections in \reffi{fi:angle}
are at the level of $5\%$ and affect the shape of the distributions at
the level of a few per cent, whereas the QCD corrections cause large
shape distortions and reach $200\%$ in the $\Delta y_{l^+ \gamma}$
and $300\%$ in the $\Delta \phi_{l^+ \gamma}$ distribution. 
The shape distortion in the $\Delta y_{l^+ \gamma}$ distribution 
originates essentially from the gluon-induced corrections, which do
not have a radiation zero.
Especially in the rapidity-difference distribution,  
effects of anomalous couplings are expected to be visible as pointed out 
in \citere{DeFlorian:2000sg}.
Similarly to the rapidity distributions discussed before,
the EW corrections are overwhelmed by QCD effects and the corresponding uncertainties
here.

\subsection{Results with anomalous triple gauge-boson couplings}
\label{se:AC}

Assuming that the SM is the low-energy limit of a more complete
theory, higher-dimensional operators can be added to the SM Lagrangian
to parametrize possible effects of new physics.  
The commonly used form of
anomalous triple gauge-boson couplings (aTGCs) goes back to
\citere{Hagiwara:1986vm} and is based on a general parametrization of the 
$\PW\PW\gamma$ and $\PW\PW\PZ$ vertices (assuming $\PW$~bosons 
coupling to conserved currents). 
For $\PW + \gamma$ production at hadron colliders, anomalous $\PW \PW \gamma$
couplings were, e.g., studied in \citeres{Baur:1993ir,DeFlorian:2000sg}.
Following these publications we assume gauge invariance as well as C and P
conservation, i.e.\ we employ the effective vertex function
\begin{align}
  \Gamma^{\mu\nu\rho}_{\PWp\PWm\gamma,\mr{AC}}\klam{q,\overline{q},p}
                                 =e &\left\{\overline{q}^{\mu}g^{\nu\rho}
                                  \klam{\Delta\kappa^{\gamma}+\la^{\gamma}\frac{q^{2}}{\MW^{2}}}
                                           -q^{\nu}g^{\mu\rho}
                                  \klam{\Delta\kappa^{\gamma}+\la^{\gamma}\frac{\overline{q}^{2}}{\MW^{2}}}
                                  \right. \nonumber \\
                                  &\left. 
                                  +\klam{\overline{q}^{\rho}-q^{\rho}}\frac{\la^{\gamma}}{\MW^{2}}
                                  \left( p^{\mu}p^{\nu} -\frac{1}{2}g^{\mu\nu}p^{2}\right)
                                  \right\} ,
\end{align}
where $\Delta\kappa$ and $\la^{\gamma}$ parametrize the strengths of
the anomalous couplings, and $e$ is the electromagnetic coupling
constant. The four-momenta of the incoming $\PW^{+}$ and $\PW^{-}$ bosons
and the photon are denoted by $q$, $\overline{q}$, and $p$,
respectively. In contrast to
\citeres{Baur:1993ir,DeFlorian:2000sg,Hagiwara:1986vm} we consider all momenta as
incoming leading to a difference 
by a global minus sign. The
anomalous couplings spoil unitarity of the S-matrix in the
limit of high energies. This problem can be bypassed by supplementing
the couplings with form factors, mimicking the onset of new physics,
which damps the effects of the aTGCs at high momentum transfer.
We use the standard choice
\begin{align}
  \Delta\kappa^{\gamma}\rightarrow \frac{\Delta\kappa^{\gamma}}
              {\left(1+\frac{M^{2}_{\PW\gamma}}{\La^{2}}\right)^{n}}\, , \qquad 
  \la^{\gamma}\rightarrow \frac{\la^{\gamma}}{\left(1+\frac{M^{2}_{\PW\gamma}}{\La^{2}}\right)^{n}}\, ,
\end{align}
where $\La$ is the scale of new physics and $M_{\PW\gamma}$ is
the invariant mass of the W-boson--photon system. The exponent $n$
is chosen in such a way 
that the form factor is sufficient to restore
unitarity. Following previous analyses we use $n=2$. 
In order to combine the contribution of the anomalous $\PW\PW\gamma$ coupling (AC)
with the NLO corrections in a consistent way, we 
extend (\ref{naive-product})
by the relative anomalous contribution $\delta_{\mr{AC}}$,
\begin{align}\label{naive-product-ac}
    \sigma^{\mr{NLO}}_{\mr{AC}}&= \sigma^{\mr{NLO\,QCD}}
                                  \klam{1+\delta_{\EW,\Pq\overline{\Pq}}+\delta_{\EW, \Pq\gamma}+\delta_{\mr{AC}}},
\end{align}
where $\delta_{\mr{AC}}$ is defined by
\begin{align}
  \delta_{\mr{AC}}&= \frac{\sigma^{\NLO\,\QCD}_{\mr{AC}}}{\sigma^{\NLO\,\QCD}}-1 \, .
\end{align}
The SM cross section $\sigma^{\NLO\,\QCD}$ is defined in (\ref{eq:sigmaqcd}),
and $\si^{\NLO\,\QCD}_{\mr{AC}}$ is the NLO QCD cross section including the aTGC contribution.
Thus, $\delta_{\mr{AC}}$ can be considered as an additional
contribution to the EW correction factor in \refeq{naive-product}. Combining the aTGCs 
in a multiplicative way with the EW corrections would require an effective-field-theory 
approach to properly account for the fact that the effective model is non-renormalizable.
In contrast, aTGCs do not conflict 
with the renormalization of QCD.
When comparing $\delta_{\mr{AC}}$ with and without a jet veto, it should be noticed
that the normalization of the two curves for $\delta_{\mr{AC}}$ is not the same,
because $\sigma^{\mr{NLO\,QCD}}$ strongly depends on the jet veto.

Experimental limits on the parameters $\Delta\kappa^{\gamma}$ and
$\la^{\gamma}$ have been updated recently in
\citeres{Aad:2013izg,Chatrchyan:2013fya,Aad:2012mr}.
We present exemplary results for
one specific point in parameter space that coincides with
the present experimental limits of \citere{Aad:2012mr},
\begin{align}
  \Delta\kappa^{\gamma}&=0.41\, ,  &\la^{\gamma}&=0.074\, ,  &\La&= 2\TeV\, . 
\end{align}
Additionally we analyse the sensitivity on the new-physics scale 
by presenting results for two more values of $\La$,
\begin{align}
    \La&=1\TeV\, ,  &\La&\rightarrow \infty\, , 
\end{align}
while $\Delta\kappa$ and $\la^{\gamma}$ are kept constant. 

Analysing the impact of aTGCs we only present results obtained with
a jet veto for the transverse-momentum distributions.
In the remaining distributions they do not differ noticeably.
Except for the transverse-mass distribution of the
charged lepton and the neutrino, we only show relative corrections
for $\La=1\TeV$ and $\La=2\TeV$ for the sake of clarity. 

\begin{figure}     
\centerline{
        \includegraphics[scale=0.6]{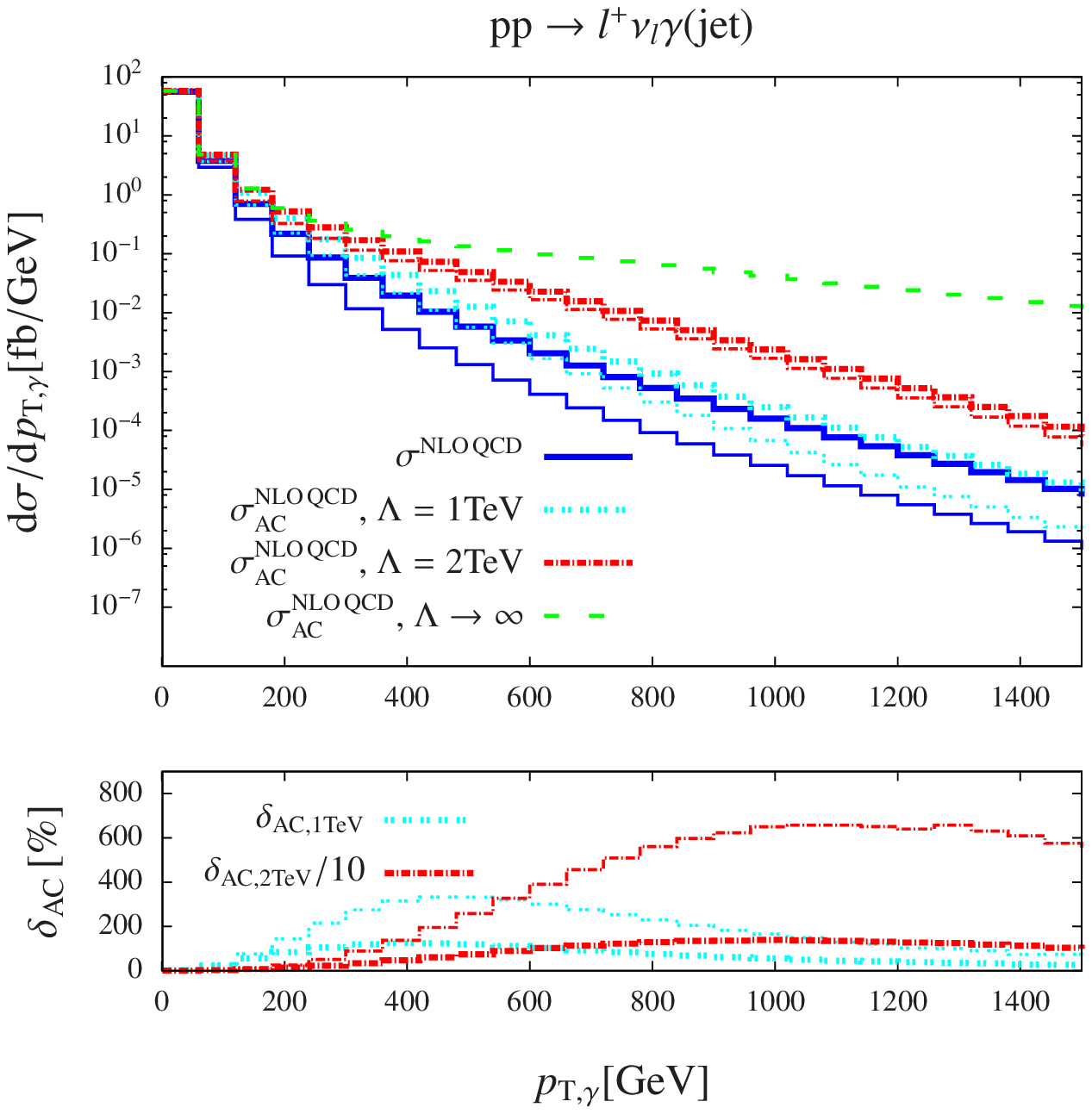}
\hspace{-1em}
        \includegraphics[scale=0.6]{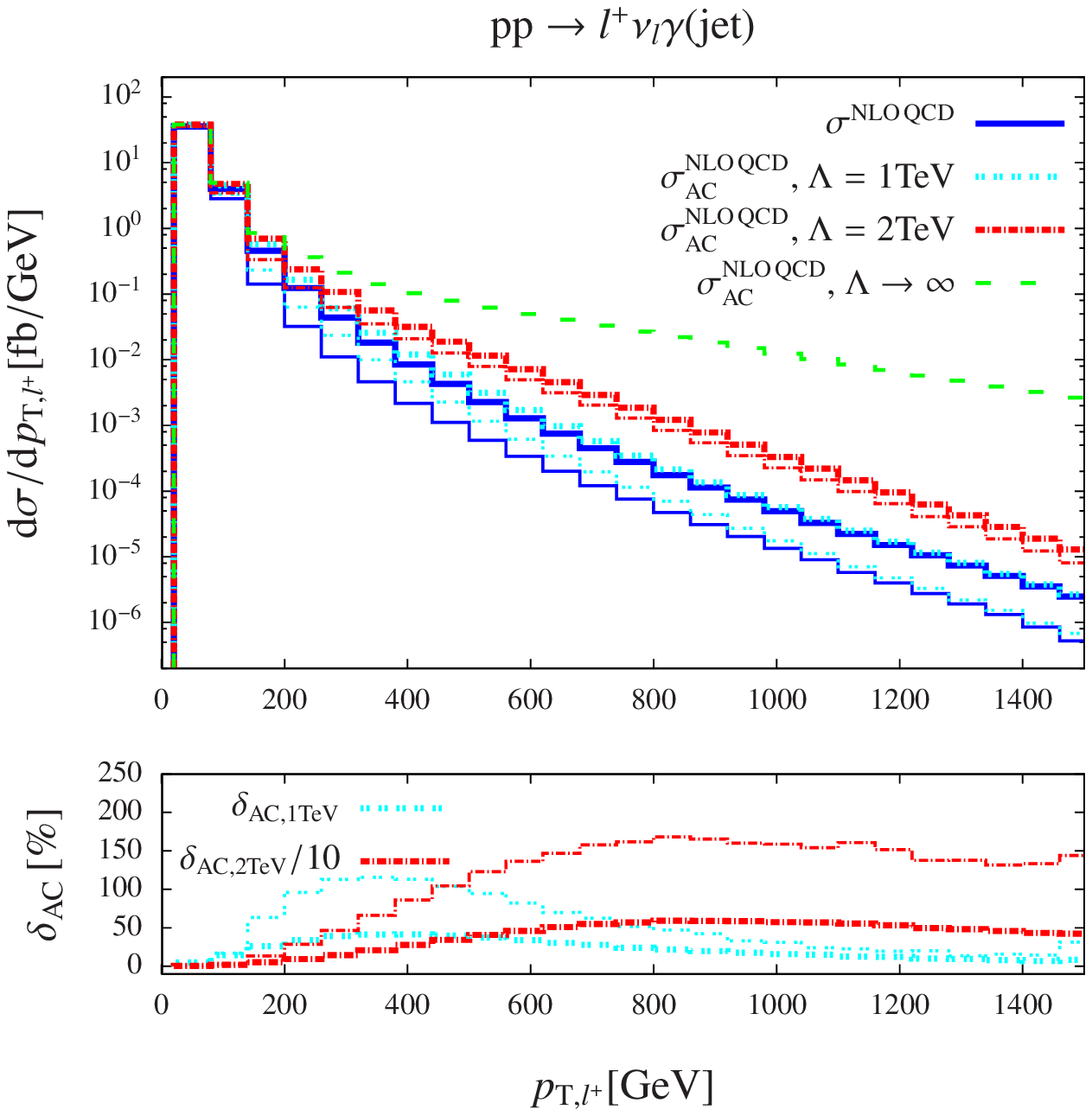}
}
\caption{\label{fi:pt_ac} Absolute and relative contributions of aTGCs
  to the transverse-momentum distributions of the photon (left) and
  the charged lepton (right). Results are presented with and without a 
  jet veto plotted with thin and thick lines, respectively. 
}
\vspace*{1cm}
\centerline{
        \includegraphics[scale=0.6]{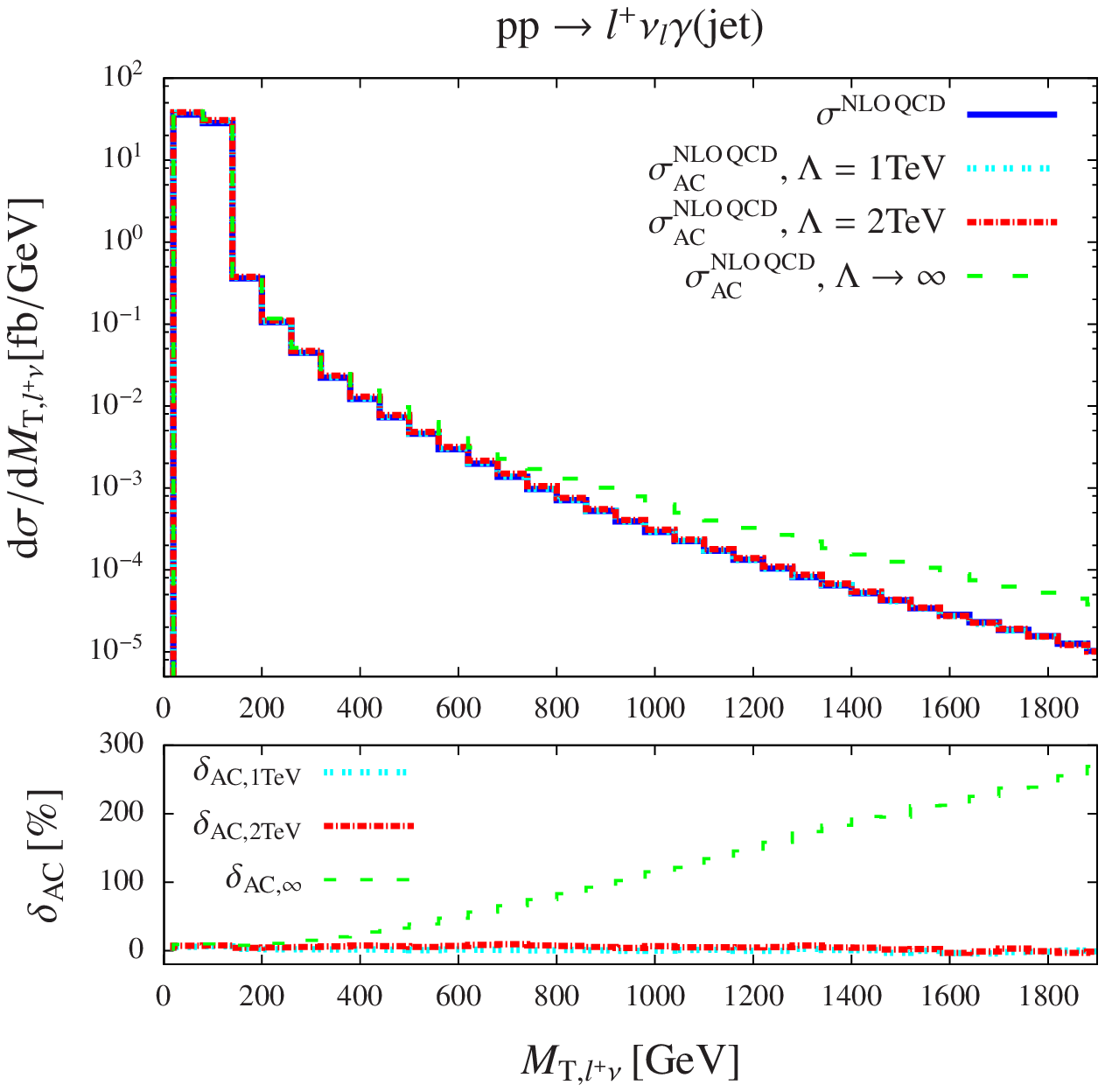}
\hspace{-1em}
        \includegraphics[scale=0.6]{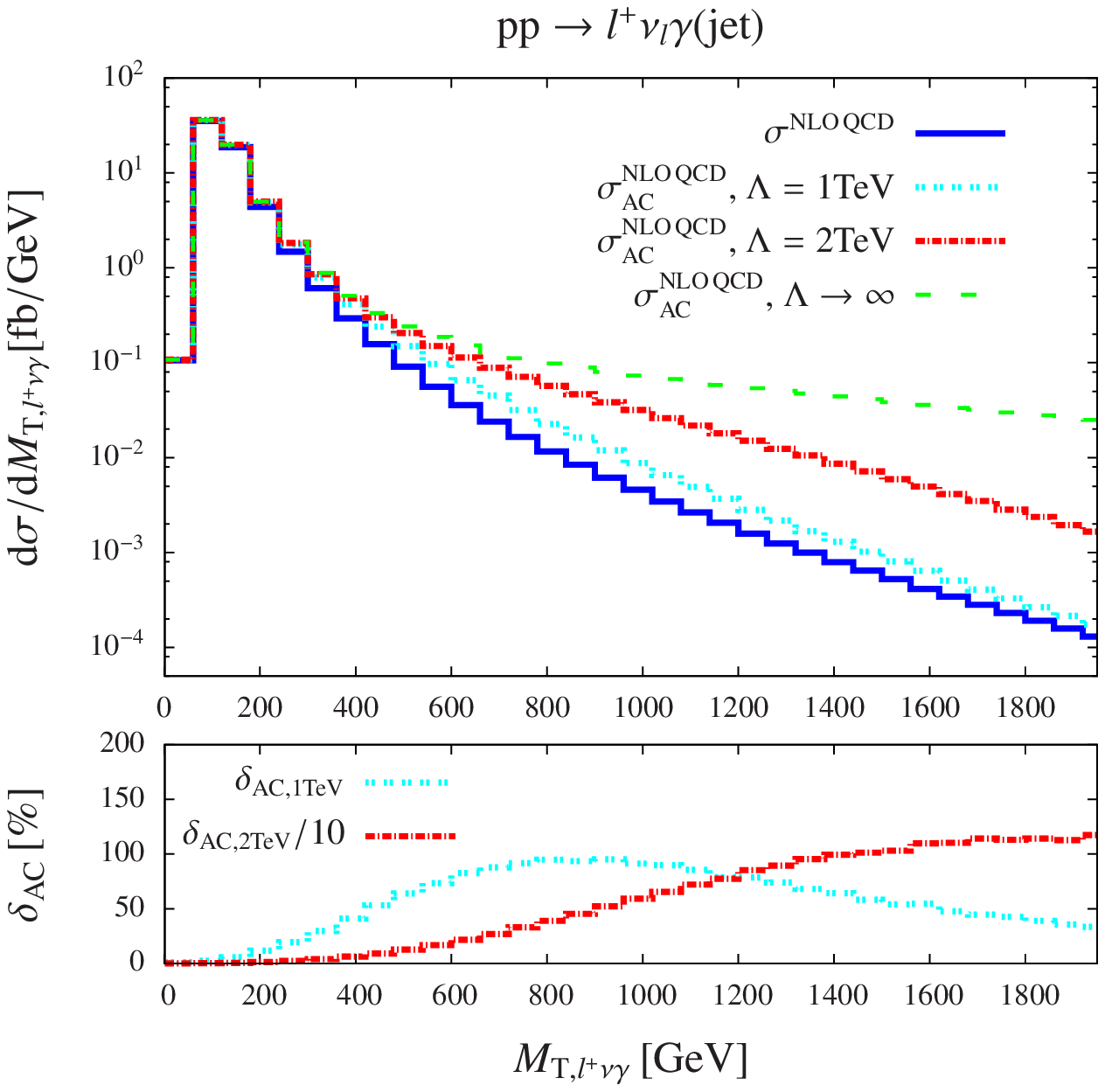}
}
\caption{\label{fi:mt_ac} Absolute and relative contributions of aTGCs
  to the transverse-mass distribution of the charged lepton and the neutrino (left)
  and to the transverse three-body mass of the charged lepton, the neutrino, and the hardest photon (right).
  }
\end{figure} 
In \reffi{fi:pt_ac} we present the transverse-momentum distributions
of the photon (left) and the charged lepton (right) including aTGCs.
In both distributions the effect of the anomalous couplings
significantly increases with growing scale $\La$. 
For $\La\to\infty$ the region 
of high transverse momenta receives huge contributions from the aTGCs.
The differential cross sections are several orders of magnitude 
larger than in the SM. The effect is less dramatic for 
$\La=2\TeV$, but still of the orders of $1000\%$ and $500\%$
for the transverse-momentum distribution of the photon 
and the charged lepton, respectively. Note that $\delta_{\mr{AC}}$ and $\delta^{\mr{veto}}_{\mr{AC}}$
are scaled down by a factor of $10$ for  $\La=2\TeV$.
Applying a jet veto increases the impact of the aTGCs considerably, since the veto has 
a large effect on the differential SM cross section, but 
a small one in the case of aTGCs. The insensitivity of the aTGCs
on the jet veto can be explained as follows: 
The aTGCs only have a large effect if the energy flow 
through the anomalous $\PW\PW\ga$
coupling is large. However, in case of a hard jet 
a substantial part of the energy is carried away by the jet.  
The picture is somewhat different for $\La=1\TeV$. In this case the 
relative contribution from the aTGCs has a maximum in the 
transverse-momentum distributions of the photon and the charged lepton 
around $500 \GeV$ and $350 \GeV$, respectively, and decreases 
for larger transverse momenta. Here the effect of the form factor, 
suppressing the impact of the aTGCs, is directly visible. 
Applying a jet veto increases the relative contribution of the 
aTGCs for the same reason as before. 

Surprisingly the impact of the aTGCs on the transverse-mass distribution 
of the charged lepton and the neutrino shown in \reffi{fi:mt_ac} (left) 
is very small and almost becomes zero for large transverse masses.
Even without a form factor the differential cross section is only
three times larger than in the SM case. 
Note that a large two-particle transverse invariant mass $\MtW$ requires
a large invariant mass $M_{\Plp\nu}$, which implies that both
intermediate W~bosons attached to the anomalous $\PW\PW\ga$ vertex are far off shell.
This observation offers an explanation for the comparably small effect of aTGCs at large
$\MtW$, which are typically driven by disturbing the unitarity
cancellations of the SM amplitude. For resonant W~bosons these cancellations
occur for longitudinally polarized W~bosons with momentum $q^\mu$
and virtuality $q^2\sim\MW^2$, where the effective W~polarization vector behaves like 
$\varepsilon_{\mathrm{L}}^\mu\sim q^\mu/\sqrt{q^2}\sim q^\mu/\MW$.
For large $\MtW$, the W~virtuality is large, $q^2\gg\MW^2$, so that 
$\varepsilon_{\mathrm{L}}^\mu\sim q^\mu/\sqrt{q^2}$ is not enhanced by a $1/\MW$
factor, and no large cancellations are necessary within the amplitude to
avoid unitarity violations in the SM. The missing $1/\MW$ enhancement in
$\varepsilon_{\mathrm{L}}$ explains the fact that the aTGC effects
are not as pronounced in the high-mass tail of $\MtW$ as compared to other
scale-dependent distributions.

In contrast, a large transverse cluster mass $\MtWA$ can be reached for outgoing on-shell
W~bosons if the photon carries away a large fraction of the momentum brought
into the $\PW\PW\ga$ vertex by the incoming W~boson, which has the
large virtuality.
This is the reason why the $\MtWA$ distribution (see \reffi{fi:mt_ac}, right)  
falls off less steeply than the $\MtW$ distribution at high scales.
Following the argument based on the leading behaviour of the effective
longitudinal polarization vector $\varepsilon_{\mathrm{L}}$ outlined above,
we expect huge aTGC effects for large $\MtWA$.
In fact,
for $\Lambda\to\infty$
the differential cross section is enhanced by roughly two orders of magnitude.
For $\La=2\TeV$ the cross section is approximately ten times larger
than the SM cross section in the region of large transverse masses.
Note that the relative contribution of the aTGCs is scaled down 
by a factor of $10$ for $\La=2\TeV$.
In case of $\La=1\TeV$ the relative anomalous contribution has a peak
around $800 \GeV$ and decreases to $30\%$ at $2\TeV$.

\begin{figure}
\centerline{
        \includegraphics[scale=0.6]{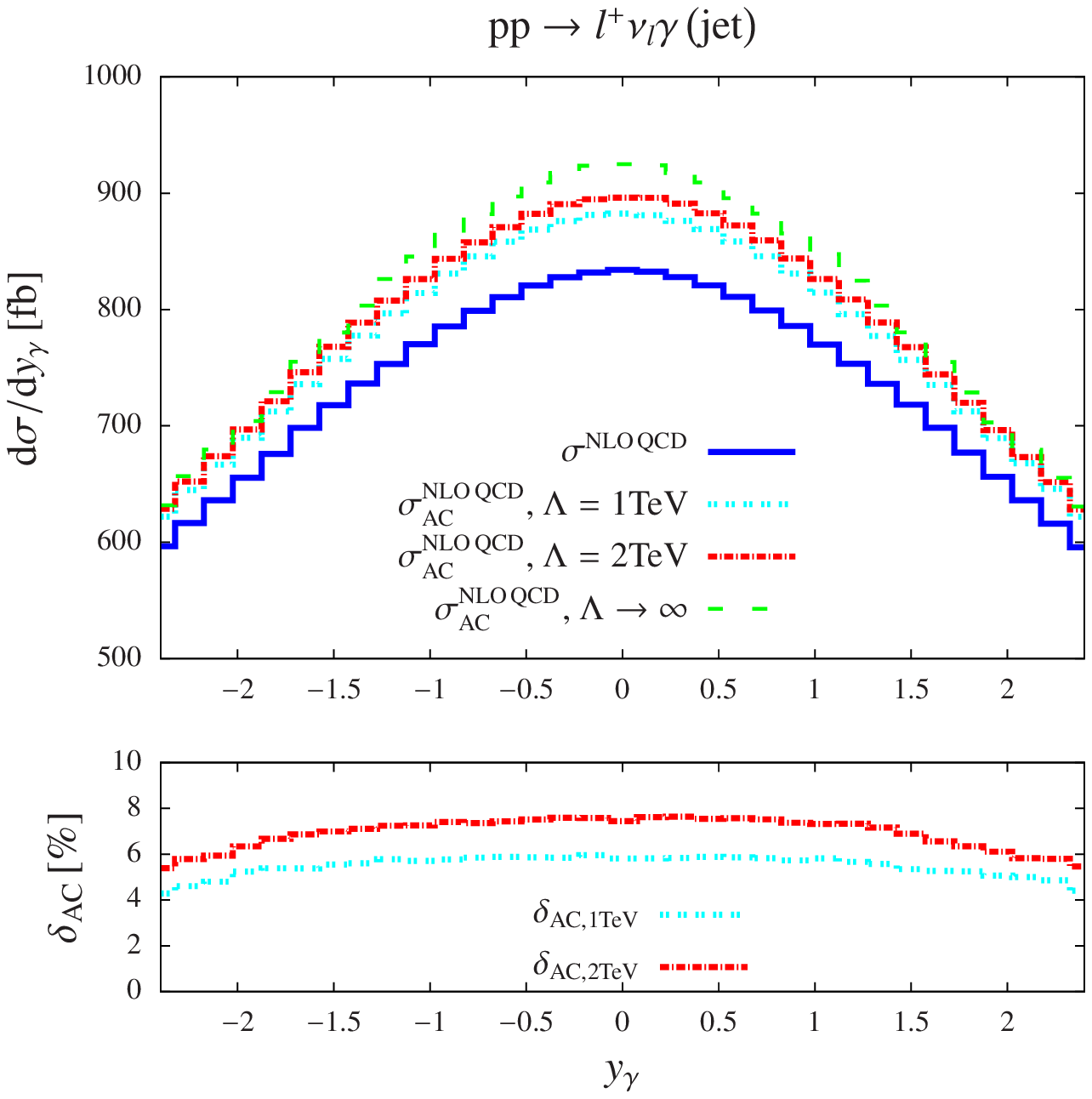}
\hspace{-1em}
        \includegraphics[scale=0.6]{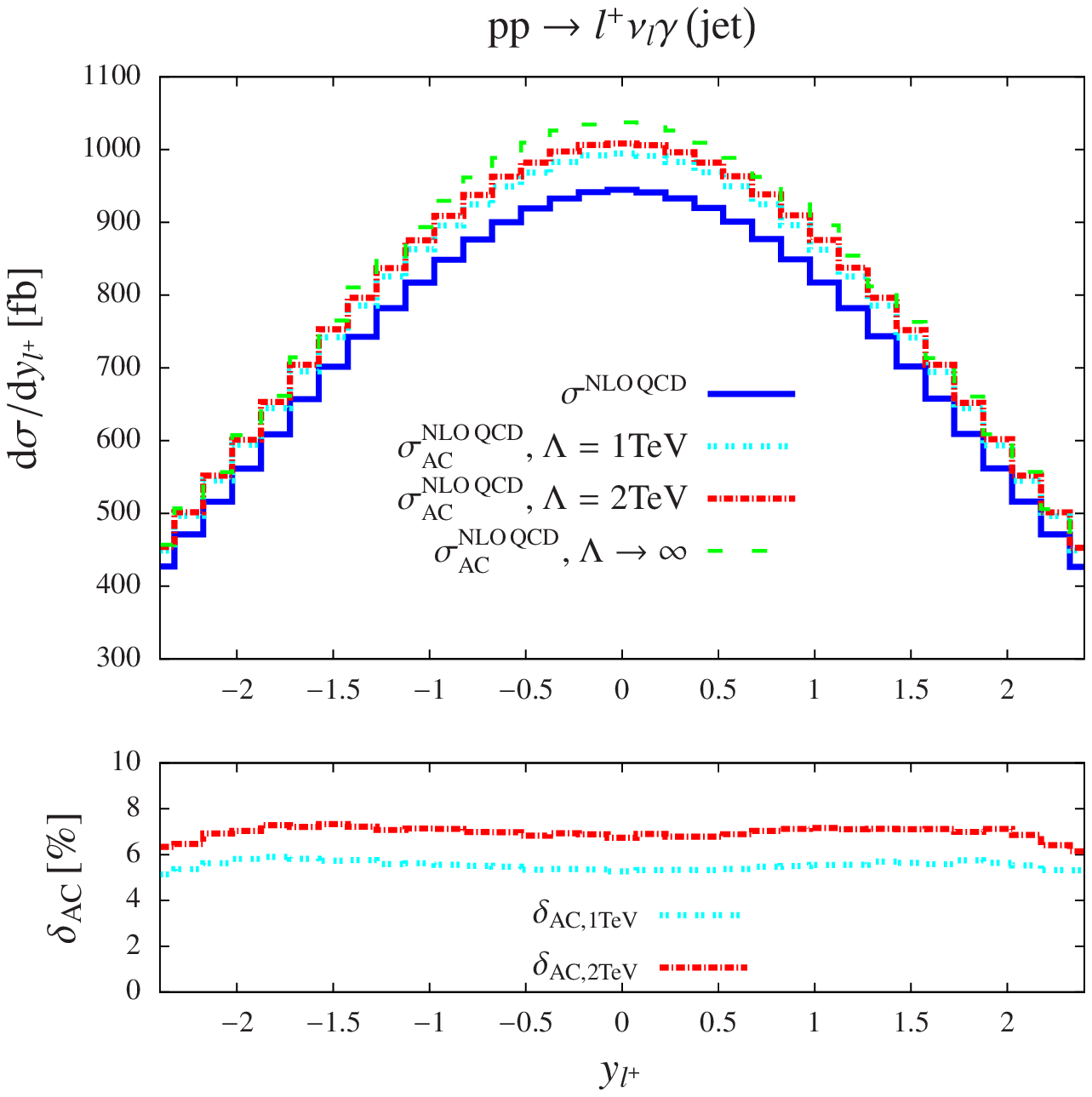}
}
\caption{\label{fi:rap_ac} Absolute and relative contributions of aTGCs
  to the rapidity distributions of the photon (left) and the charged
  lepton (right).}
\vspace*{1cm}
\centerline{
        \includegraphics[scale=0.6]{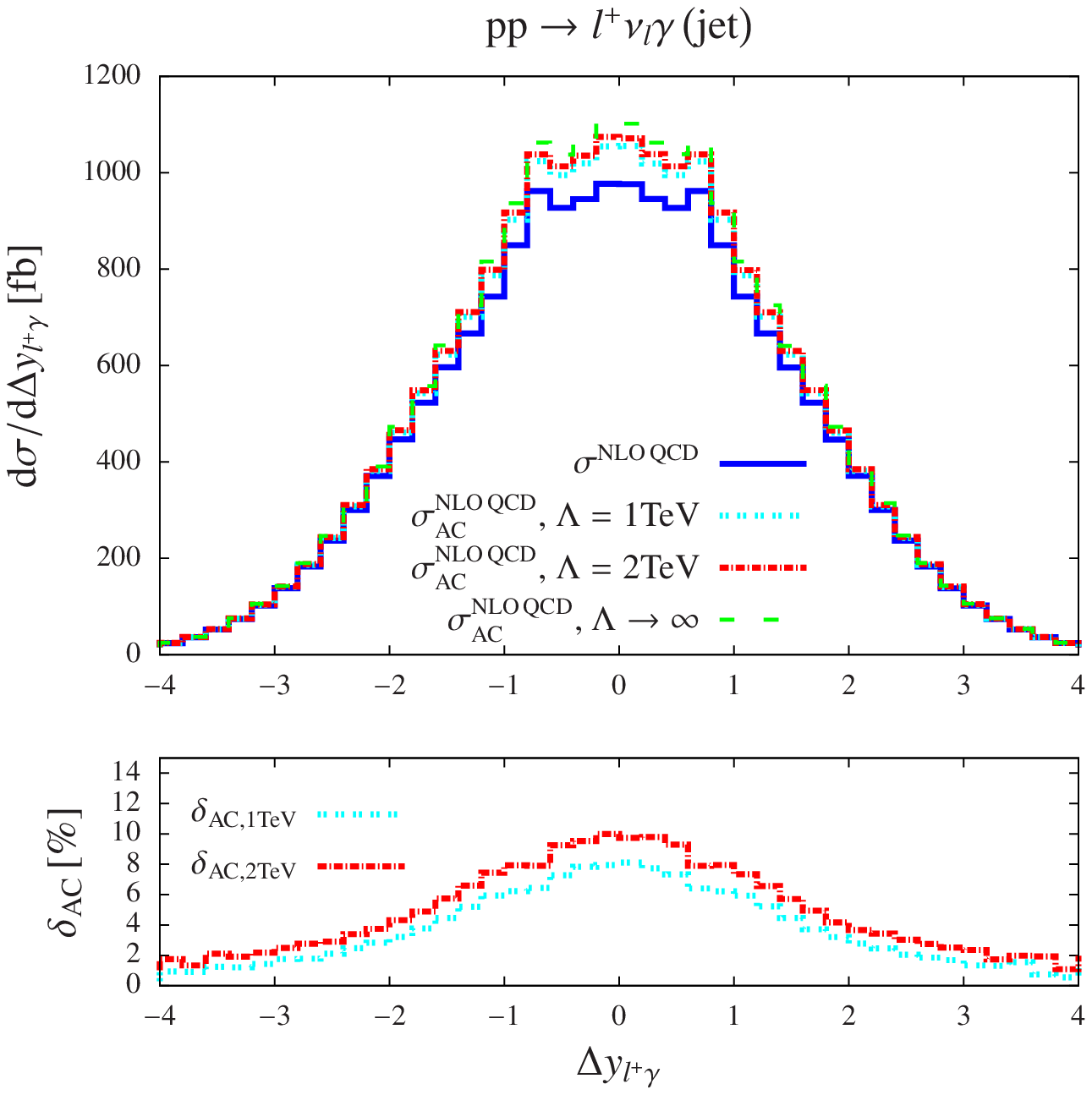}
\hspace{-1em}
        \includegraphics[scale=0.6]{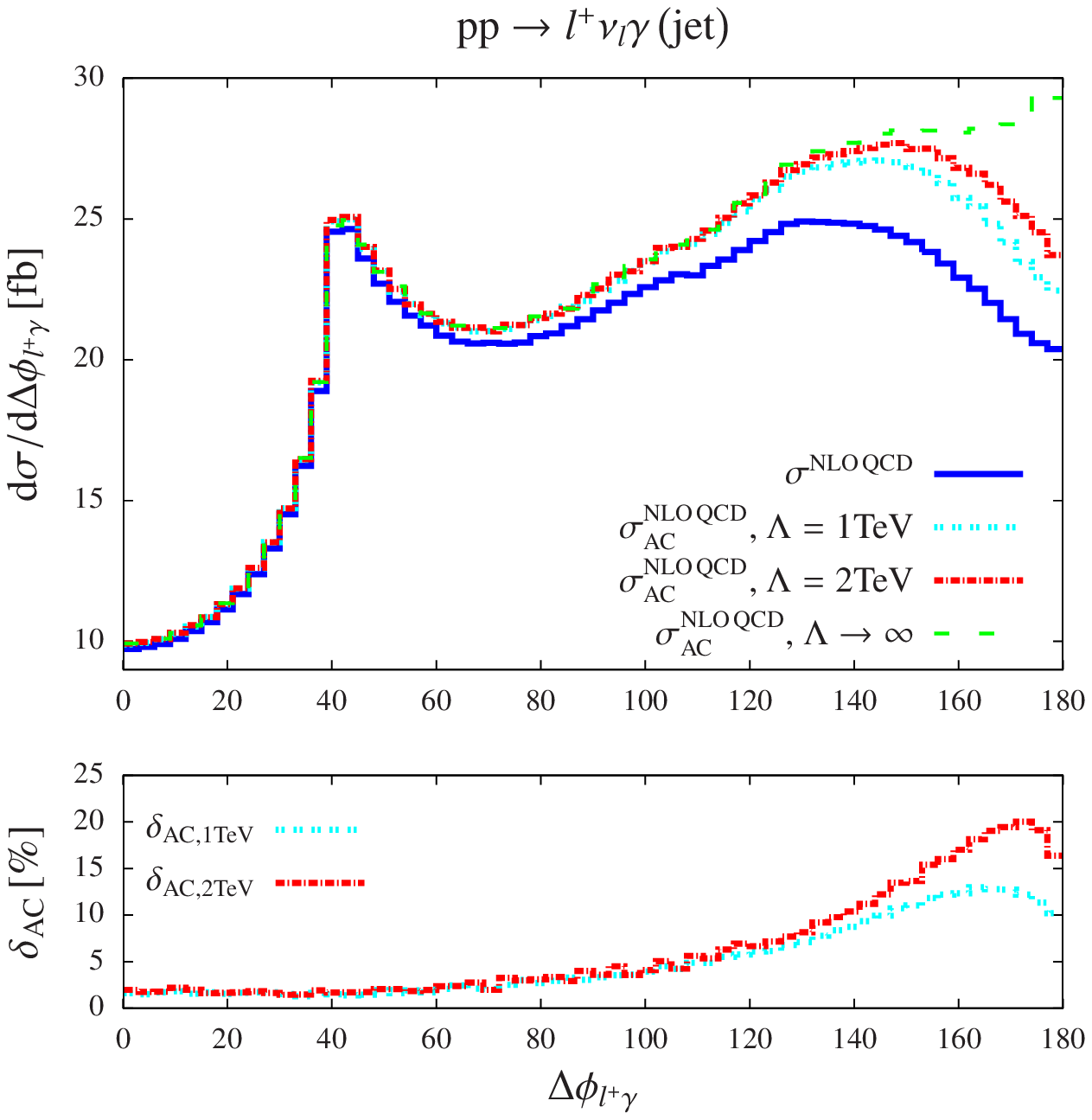}
}
\caption{\label{fi:angle_ac} Absolute and relative contributions of aTGCs
  to the rapidity-difference distribution (left) and the
  azimuthal-angle-difference distribution (right) between the photon
  and the charged lepton.}
\end{figure}

The impact of the anomalous couplings on the rapidity distributions 
presented in \reffi{fi:rap_ac}
amounts to about $5\%$ and $7\%$ for $\La=1\TeV$ and $\La=2\TeV$,
respectively.  In contrast to the rapidity distributions, the rapidity
difference between the charged lepton and the photon shown in
\reffi{fi:angle_ac} (left) receives a significant shape distortion.
The contribution of the aTGCs is similar for the two values of the
scale $\La$ and reaches $10\%$ around $\Delta y_{\Pl^{+}\gamma}=0$. This
can be explained by the radiation zero that appears for a stable W
boson in $\PW + \gamma$ production at $\Delta y_{\Pl^{+}\gamma}=0$.
This effect is, however, washed out, because the W boson decays and only the
charged lepton can be detected.
Additionally at NLO, QCD radiation fills the radiation zero.
Since the contributions of anomalous couplings do not exhibit a
radiation zero they lead to  a sizeable enhancement compared to the SM
prediction around $\Delta y_{\Pl^{+}\gamma}=0$.

Focusing on the angular difference between the charged lepton and the
photon (\reffi{fi:angle_ac}, right) we find a similar shape distortion 
for the two scales $\La=1\TeV$ and $\La=2\TeV$. 
The correction induced by
the aTGCs has a maximum if the angle between the photon and the
charged lepton is almost $180^{\circ}$. The effect is even larger 
if the scale goes to infinity. 
Since the effect of aTGCs is enhanced at high energies,
the LHC running at $14\TeV$ will allow to set very tight limits.

\section{Conclusions}
\label{se:concl}

The production of $\PW + \gamma$ final states at hadron colliders
represents the ideal process to investigate the interaction of W~bosons with photons
at high energies. Deviations from the standard form of the 
$\PW\PW\gamma$ interaction vertex, as typically predicted by Standard Model extensions,
are quantified in terms of anomalous couplings which are already experimentally
constrained by the analysis of W-pair production at LEP and by $\PW + \gamma$
production 
at the Tevatron. The run~2 phase of the LHC, starting in 2015, will tighten
these constraints significantly, rendering predictions for $\PW + \gamma$
production at the few-per-cent level necessary.

In this paper we have improved the state-of-the-art knowledge of $\PW + \gamma$
production on the side of electroweak higher-order corrections.
Specifically, we have calculated the full next-to-leading-order electroweak
corrections to $\PW + \gamma$ production with leptonically decaying W~bosons, 
taking into account all off-shell effects of the W~boson using the
complex-mass scheme and including 
effects originating from initial-state photons.
For a phenomenologically sound definition of the $\PW + \gamma$ signature,
it is necessary to consistently separate hard photons from jets.
To this end, we employ a quark-to-photon fragmentation function \'a la Glover and
Morgan.

While electroweak corrections to integrated cross sections turn out to be
at the level of few per cent, in line with previous predictions, they grow
to several $10\%$ in distributions where high energy scales matter.
Moreover, in the high-momentum tail of the
transverse-momentum distribution of the hard photon we
observe a huge impact of the photon-induced channels, which inherit large
uncertainties from the photon PDF at large $x$. 
In order to bring uncertainties down to the $\sim10\%$ level (or better)
in this regime,
phenomenological improvements on the photon PDF will be necessary.
To some extent, a jet veto, which excludes overwhelming QCD corrections
originating from hard jet emission, reduces the large size of the photon-induced
channels as well.
Generically, distributions in angles and rapidities receive only small 
uniformly distributed electroweak corrections, which are shadowed
by QCD effects.

We have reproduced the next-to-leading-order QCD corrections as well
and discussed the effects of anomalous triple gauge-boson couplings on
various NLO-QCD-corrected distributions.  
For full state-of-the-art
predictions the new results on electroweak corrections should be
combined with the next-to-next-to-leading-order QCD corrections which
have been recently presented in the literature.  This combination
should provide the necessary precision in predictions required for the
coming data analysis at the LHC at its design energy and luminosity.

\subsection*{Acknowledgements}
This project is supported by the German Research Foundation (DFG) via
grant DI 784/2-1 and the research training groups 
GRK 1102 ``Physics at Hadron Colliders'' and
GRK 1147 ``Theoretical Astrophysics and Particle Physics''.

\bibliography{Bibliography}

\end{document}